\documentclass[%
 reprint,
superscriptaddress,
nofootinbib,
 amsmath,amssymb,
 aps,
prd,
]{revtex4-2}

\usepackage{graphicx}%
\usepackage{dcolumn}%
\usepackage{bm}%

\usepackage[english]{babel}
\usepackage{fancyhdr}
\usepackage{amsmath}
\usepackage{amssymb}
\usepackage{amsfonts}
\usepackage{psfrag}
\usepackage[applemac]{inputenc}
\usepackage[bf,footnotesize,justification=Justified,format=plain]{caption}
\usepackage[dvipsnames]{xcolor}
\usepackage[colorlinks=True, citecolor=blue, linkcolor=blue, urlcolor=blue,linktocpage]{hyperref}
\usepackage{amsthm}
\usepackage{gensymb}
\usepackage{subfig}
\usepackage{physics}
\usepackage{array}
\usepackage{tcolorbox, mathtools}
\usepackage{soul}
\usepackage{aas_macros}
\usepackage[normalem]{ulem}

\newcommand{\sch}{Schwarzschild }

\newcommand\underrel[2]{\mathrel{\mathop{#2}\limits_{#1}}}
\newcommand{\cb}{ }

\begin{document}

\title{Quadratic Quasi-Normal Modes of a \sch Black Hole}%

\author{Bruno Bucciotti}\email{bruno.bucciotti@sns.it}
\author{Leonardo Juliano}\email{leonardo.juliano@sns.it}

\affiliation{%
 Scuola Normale Superiore, Piazza dei Cavalieri 7, 56126, Pisa, Italy
}%
\affiliation{INFN Sezione di Pisa, Largo Pontecorvo 3, 56127 Pisa, Italy}
\author{Adrien Kuntz}
 \email{adrien.kuntz@sissa.it}
\affiliation{SISSA, Via Bonomea 265, 34136 Trieste, Italy and INFN Sezione di Trieste}
\affiliation{IFPU - Institute for Fundamental Physics of the Universe, Via Beirut 2, 34014 Trieste, Italy}
\author{Enrico Trincherini}\email{enrico.trincherini@sns.it}
\affiliation{%
 Scuola Normale Superiore, Piazza dei Cavalieri 7, 56126, Pisa, Italy
}%
\affiliation{INFN Sezione di Pisa, Largo Pontecorvo 3, 56127 Pisa, Italy}

\date{\today}%

\begin{abstract}
Quadratic quasi-normal modes, generated at second order in black hole perturbation theory, are a promising target for testing gravity in the nonlinear regime with next-generation gravitational wave detectors.
While their frequencies have long been known, their amplitudes remain poorly studied. We introduce regular variables and compute amplitudes for \sch black holes with the Leaver algorithm. We find a nonlinear ratio $\mathcal{R}\simeq0.154e^{-0.068i}$ for the most excited $\ell=4$ mode, matching results from Numerical Relativity. We also predict new low-frequency $\ell=2$ quadratic modes.
\end{abstract}
\maketitle

\textbf{Introduction}
Observations of binary black hole (BH) mergers hold promise for testing General Relativity (GR) in the strong gravity regime with unprecedented precision and perhaps even finding evidence of novel phenomena in the gravitational domain.  While the precise moment of the actual merger can only be described using Numerical Relativity (NR), analytic perturbation theory plays a fundamental role at earlier and later times. In this letter, we will focus on the final phase of the merger process, known as the ringdown, specifically examining the scenario where the black hole formed in the process has zero angular momentum. 

Linear perturbations around the spacetime geometry resulting from the merger are described {\cb at intermediate times} by a superposition of a discrete set of damped sinusoids, known as quasinormal modes (QNMs)~\cite{Berti_2009}. They are characterized by complex frequencies determined in GR solely by the mass and angular momentum of the black hole. At linear order and for zero BH spin, the dynamics of the two physical gravitational wave polarizations can be described by two ``master scalar" variables $ \Psi_{\pm}^{(1)}$ that obey independent wave equations in a potential $V_{\pm}$:
\begin{equation}
\Big( \Box - V_\pm \Big) \Psi_\pm^{(1)}  = 0 \; .
\end{equation}
The equations are decoupled because the two perturbations transform oppositely under parity (hence the subscript $\pm$), which is a symmetry of the Schwarzschild geometry. The resulting frequencies $\omega_{\ell m n}$, labeled by two angular harmonic numbers $(\ell, m)$ and an overtone number $n$, can be utilized for accurate gravity testing~\cite{Dreyer:2003bv, Berti_2006,Berti_2016, Bhagwat:2016ntk,Yang:2017zxs, Baibhav:2017jhs, Brito:2018rfr, Baibhav:2018rfk, Bhagwat:2019dtm, Maselli_2020, CalderonBustillo:2020rmh, Capano:2020dix, JimenezForteza:2020cve, Ota:2021ypb, Maselli:2023khq, Baibhav:2023clw,2019PhRvD..99j4077C,2019PhRvD.100d4061M,Franchini:2022axs,Silva:2024ffz,Ghosh:2021mrv,LIGOScientific:2016lio,LIGOScientific:2020tif,Franchini:2023eda,Toubiana:2023cwr,Hirano:2024fgp}. On the other hand, the amplitudes of these modes depend on the initial conditions of the merger and should be fitted against data or Numerical Relativity simulations~\cite{Ghosh:2021mrv,LIGOScientific:2016lio,LIGOScientific:2020tif,Buonanno:2006ui,Berti:2007fi,Berti:2007zu,Baibhav:2017jhs,Giesler:2019uxc,Cheung:2023vki}. 

Since the perturbations decay exponentially, it is not surprising that linear Black Hole Perturbation Theory (BHPT)~\cite{Regge:1957td,Zerilli:1970aa} can effectively model the ringdown after a time roughly equivalent to the QNM decay timescale. However, during the initial phase of the ringdown, just after the peak of the strain, it becomes essential to account for nonlinear effects. This is particularly relevant because the first few cycles encompass most of the observable signal, providing the best opportunities for precise measurements. Indeed, recent studies have shown that second-order BHPT may be necessary to accurately describe the signals obtained from NR~\cite{Ma_2022,London_2014,Mitman:2022qdl, Cheung:2022rbm,Cheung:2023vki,Khera:2023oyf,Zhu:2024rej,Redondo-Yuste:2023seq}.

To second order in perturbation theory, the dynamics is governed by the same wave equation for the second order master scalars $ \Psi_\pm^{(2)}$, but now with a source term  $S^{(2)}$ quadratic in the linear perturbations,
\begin{equation}\label{Eq:secondorderEq}
\Big( \Box - V_\pm\Big) \Psi_\pm^{(2)} = S^{(2)} \; .
\end{equation}
As a consequence, there exists a set of quadratic QNMs (QQNMs) generated by the product of two linear modes.  For each $(\ell, m)$ the waveform contains additional modes whose frequencies can be trivially obtained from the product of two monochromatic first-order waves as $\omega^\mathrm{Q} = \omega_{\ell_1 m_1 n_1} + \omega_{\ell_2 m_2 n_2}$ or $\omega^\mathrm{Q} = \omega_{\ell_1 m_1 n_1} - (\omega_{\ell_2 m_2 n_2})^*$~\cite{Lagos_2023}, where $*$ denotes complex conjugation.
A crucial feature of QQNMs is that, unlike the linear ones, their {\it amplitude} is also entirely determined by GR. This essential property arises from the fact that these modes solve a non-homogeneous equation~\eqref{Eq:secondorderEq}. Despite this fact, to date, only a few specific second-order amplitudes have been computed, either by fitting NR simulations~\cite{Ma_2022,London_2014,Mitman:2022qdl, Cheung:2022rbm,Cheung:2023vki,Khera:2023oyf,Zhu:2024rej} or by integrating BHPT equations~\cite{Nakano:2007cj,Ioka:2007ak,Bucciotti:2023ets,Redondo-Yuste:2023seq,Ma:2024qcv,Perrone:2023jzq}. Since at least the loudest QQNM will be observable by next-generation interferometers~\cite{Yi:2024elj}, there is an increased need for precise computation of these amplitudes. 

This letter aims to provide the amplitude of all low multipole QQNMs for a Schwarzschild BH. 
One of the main obstacles in solving the second-order equation can be traced back to the redundancy that inevitably arises when describing the two degrees of freedom (DOF) of a massless spin-2 particle in terms of the metric tensor. Conceptually, the procedure should follow what is done at linear order and appears straightforward: one first fixes the gauge redundancy associated with diffeomorphism invariance. Secondly, one selects two master scalar variables $ \Psi_\pm^{(2)}$, expressed in terms of the metric components, which capture the dynamics of the physical DOF. Once the equation for each $\Psi_\pm^{(2)}$ is solved with the appropriate boundary conditions, it becomes possible to reconstruct the full metric and transition to an outgoing radiation gauge, where the amplitude of the physical gravitational wave observable at infinity can be computed.

Regrettably, as noted in the literature~\cite{Gleiser:1995gx,Nicasio_1999,Gleiser_2000,Brizuela:2006ne,Brizuela:2007zza,Brizuela:2009qd,Nakano:2007cj,Ioka:2007ak,Ma:2024qcv}, when the choice of a second-order master scalar is equivalent to the first-order one, quadratic perturbations exhibit a divergent behavior at infinity or the black hole horizon. Although this divergence will eventually become immaterial, it signifies a poor choice of variables, making the extraction of the subleading physical effect extremely challenging. To circumvent this problem, we will demonstrate how to define two new master scalars for arbitrary $(\ell, m)$ that are regular everywhere. With these variables, the resulting equations, along with appropriate QNM boundary conditions, can be integrated using the Leaver algorithm~\cite{leaver}. 
While we explicitly present in this article a set of amplitudes for the first few QQNMs, our complete results are available online~\cite{csvQuadratic}.
In the following, we will use units in which $G=c=1$.

\textbf{Second-order BHPT}
\begin{figure*}
\includegraphics[width=\textwidth]{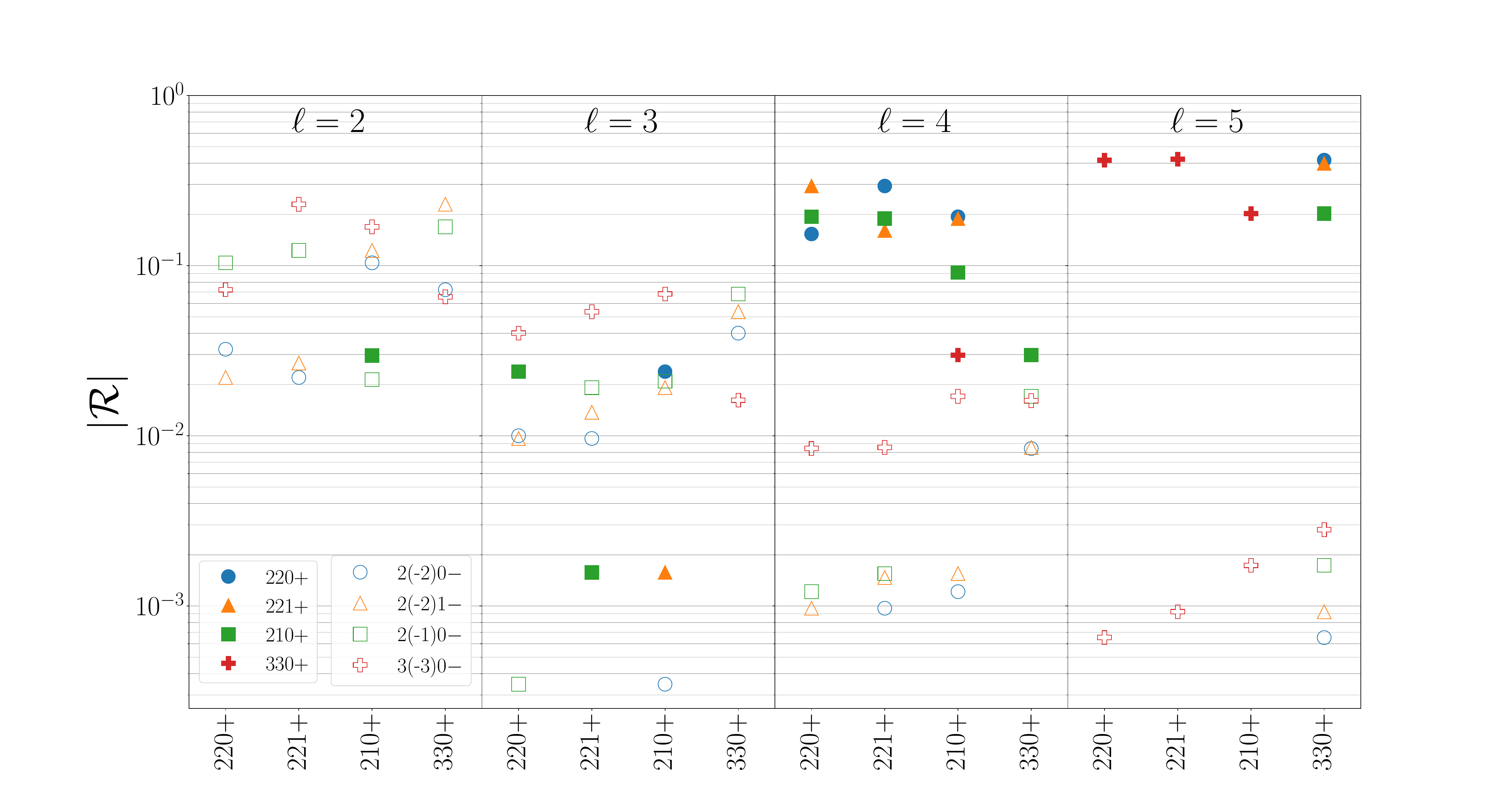}
\caption{\label{fig:ampl}Nonlinear ratio of amplitudes  $|\mathcal{R}|$ for different mode numbers. The $x$ axis represents the mode numbers $\ell m n \mathfrak m$ of the first linear parent mode, while the second linear parent mode is represented with different markers. Filled symbols represents the combination of two normal modes with $m>0$ ($\omega^\mathrm{Q} = \omega_{\ell_1 n_1 +} + \omega_{\ell_2 n_2 +}$), while empty ones represents a normal mode with $m>0$ combined with a mirror mode with $m<0$ ($\omega^\mathrm{Q} = \omega_{\ell_1 n_1 +} - \big(\omega_{\ell_2 n_2 +} \big)^*$). A missing marker means that $|\mathcal{R}|<2.5 \times 10^{-4}$. {\cb Notice that most points in this figure correspond to contributions to various spherical harmonics, as each point corresponds to its
own $m = m_1+m_2$ mode and $m_1$, $m_2$ vary.}}
\end{figure*}
We expand the spacetime metric as $g_{\mu \nu} = \bar g_{\mu \nu} + \varepsilon h^{(1)}_{\mu \nu} + \varepsilon^2 h^{(2)}_{\mu \nu}$, where $\varepsilon$ is a small bookkeeping parameter for the perturbation order -- physically corresponding to the small magnitude of linear QNMs (LQNMs) -- and $\bar g_{\mu \nu} \mathrm{d}x^\mu \mathrm{d}x^\nu = - f(r) \mathrm{d}t^2 + f(r)^{-1} \mathrm{d}r^2 + r^2 \big( \mathrm{d}\theta^2 + \sin^2 \theta \mathrm{d}\phi^2 \big)$ is the \sch metric, with $f(r) = 1-2M/r$. Up to second order, Einstein's equations read
\begin{align}
    G^{(1)}_{\mu \nu} \big[ h^{(1)} \big] &= 0 \label{eq:EinstEqFirstOrder} \; ,\\
    G^{(1)}_{\mu \nu} \big[ h^{(2)} \big] &= -  G^{(2)}_{\mu \nu} \big[ h^{(1)}, h^{(1)} \big] \label{eq:EinstEqSecondOrder} \; ,
\end{align}
where we schematically denoted by $G^{(1)}_{\mu \nu} \left[ \cdot \right]$ the part of the Einstein tensor linear in perturbations and by $G^{(2)}_{\mu \nu} \big[ \cdot, \cdot \big]$ its part bilinear in perturbations, suppressing indices on $h_{\mu \nu}$ for clarity. Thus, the LQNM $h^{(1)}$ will provide a source term for the QQNM $h^{(2)}$; notice that the same differential operator acts similarly on $h^{(1)}$ and $h^{(2)}$ on the left-hand side of these equations. Spherical symmetry of the background allows us to split spacetime between a ``$t-r$ plane'' $\mathcal M$ (indices $a,b,\dots$ and metric $\bar g_{ab} = \mathrm{diag}\left(-f,f^{-1} \right)$) and the 2-sphere $\mathcal{S}$ (indices $A,B,\dots$ and metric $\Omega_{AB} = \mathrm{diag}(1, \sin^2 \theta)$). We decompose metric perturbations according to
\begin{align}
    h_{ab}^{(i)} &= \sum_{\omega, \ell, m} e^{-i \omega t} h_{ab, \ell m\omega}^{(i)} Y^{\ell m} \label{eq:hab} \; ,\\
    h_{aB}^{(i)} &= \sum_{\omega, \ell, m} e^{-i \omega t} \big[ h_{a+, \ell m\omega}^{(i)} Y_B^{\ell m} +  h_{a-, \ell m\omega}^{(i)} X_B^{\ell m} \big] \label{eq:haB} \; ,\\
    h_{AB}^{(i)} &= \sum_{\omega, \ell, m} e^{-i \omega t} \big[ h_{\circ, \ell m\omega}^{(i)} \Omega_{AB} Y^{\ell m} \nonumber \\
    &+ h_{+, \ell m\omega}^{(i)} Y_{AB}^{\ell m} +  h_{-, \ell m\omega}^{(i)} X_{AB}^{\ell m}  \big] \label{eq:hAB} \; ,
\end{align}
where $i=1,2$ is the order of perturbation, the sum goes over all QNM frequencies $\omega$ and mode numbers $\ell, m$, and $Y, Y_A, X_A, Y_{AB}, X_{AB}$ are a set of tensor spherical harmonics defined in app.~\ref{app:technical}.
To streamline the discussion, we will omit the indices $\ell m\omega$ from variables in the following. We focus on perturbations with $\ell \geq 2$, as according to the peeling theorem~\cite{1961RSPSA.264..309S,1962RSPSA.270..103S,1962JMP.....3..566N}, these are the only ones carrying radiation to infinity. Following the approach outlined in~\cite{Spiers:2023mor}, we define a gauge-invariant variable $\tilde h_{\mu \nu}$, decomposed in spherical harmonics as in~\eqref{eq:hab}-\eqref{eq:hAB}, which coincides with the metric perturbation $h_{\mu \nu}$ in Regge-Wheeler-Zerilli (RWZ) gauge where $h_\pm = h_{a+} = 0$. Detailed expressions are provided in app.~\ref{app:technical}. 
On the other hand, we want to extract QNM amplitudes at large distances and in the physical Transverse-Traceless (TT) gauge, where the (real) $\mathfrak h_+$ and $\mathfrak h_\times$ components of the metric perturbation $h_{\mu \nu} = h_{\mu \nu}^{(1)} + \varepsilon h_{\mu \nu}^{(2)}$ are, up to $\mathcal{O}(r^{-2})$:
\begin{equation}\label{eq:QNMTT}
     \mathfrak h_+ - i \mathfrak h_\times = \frac{ M}{r} \sum_{\ell m \mathcal{N}} \mathcal{A}_{\ell m \mathcal{N}} e^{i \omega_{\ell \mathcal{N}} (r_*-t)} \vphantom{|}_{-2}Y^{\ell m}(\theta, \phi) \; .
\end{equation}
In this equation, which applies to both linear and quadratic order, $\vphantom{e}_{-2}Y$ are the spin-weighted spherical harmonics, $r_* = r + 2M \log(r/2M-1)$ is the tortoise coordinate, $\mathcal{A}_{\ell m \mathcal{N}}$ and $\omega_{\ell \mathcal{N}}$ are the amplitudes and frequencies of QNMs, %
and $\mathcal{N}$ is an additional mode number defined as follows.
For linear modes, $\mathcal{N} = (n, \mathfrak m)$ is composed of the overtone number $n$ and of the mirror modes $\mathfrak m = \pm$ . A mirror mode has a frequency related to regular modes by complex conjugation:  $\omega_{\ell n -} = - \omega_{\ell n +}^*$. Note that for \sch BHs, QNMs frequencies do not depend on the mode number $m$; we will use the convention that regular $\mathfrak m =+$ frequencies have positive real part.
For quadratic modes, $\mathcal{N} = (\ell_1, m_1, n_1, \mathfrak m_1) \times (\ell_2, m_2, n_2, \mathfrak m_2)$ contains the mode numbers of the product of linear modes which generate them~\cite{Lagos_2023}. 
Finally, we assume (as is common in the literature~\cite{Berti:2007fi,Berti:2007zu,Yi:2024elj,Isi:2021iql}) that the strain~\eqref{eq:QNMTT} {\cb enjoys an additional equatorial symmetry}. This implies that mirror modes with $m<0$ have the same amplitude as regular modes with $m'=-m>0$ up to a phase. This assumption is typically well satisfied in practise~\cite{Berti:2007fi}~\footnote{Notice that mirror modes are related to but different than retrograde modes introduced in~\cite{Isi:2021iql,Cheung:2023vki}, {\cb which are modes with $\mathrm{sign}( \text{Re} \, \omega m) <0$}, and whose amplitude are suppressed in a typical merger~\cite{Berti_2006,Buonanno:2006ui,London_2014,Lim:2019xrb}.}, and it fixes the parity of any mode to $(-1)^{\ell+m}$.

By relating $\tilde h$ to $\mathfrak h_+$ and $\mathfrak h_\times$, it can be checked that $\tilde h_{ab}^{(i)} = \mathcal{O}(r)$, $\tilde h_{a-}^{(i)} = \mathcal{O}(r)$, $\tilde h_{\circ}^{(i)} = \mathcal{O}(r^2)$ for $r \rightarrow \infty$ regardless of the gauge choice (since $\tilde h$ is gauge-invariant). The precise relationships between asymptotic amplitudes of the various variables {\cb are given in the GitHub package~\cite{csvQuadratic} and in our companion paper~\cite{Bucciotti:2024jrv}}. We now define modified RWZ variables as
\begin{align} \label{eq:RWvar}
    \Psi_\mathrm{-}^{(i)} &= \frac{r}{\lambda} \bigg[ D_r \tilde h_{t-}^{(i)} - D_t \tilde h_{r-}^{(i)} - \frac{2}{r} \tilde h_{t-}^{(i)} \bigg] + g^{(i)}_\mathrm{-}(r) \; , \\
    \Psi_\mathrm{+}^{(i)} &= \frac{2r}{\ell(\ell+1)} \Bigg[ \frac{\tilde h_\circ^{(i)}}{r^2} + \frac{1}{\lambda + 3M/r} \Bigg( f^2(r)\tilde h_{rr}^{(i)} \nonumber \\
    & - r f(r) D_r \bigg( \frac{\tilde h_\circ^{(i)}}{r^2} \bigg) \Bigg) \Bigg] + g^{(i)}_\mathrm{+}(r) \label{eq:Zvar} \; ,
\end{align}
where $D$ denotes the covariant derivative on $\mathcal{M}$, $\lambda = (\ell+2)(\ell-1)/2$, and $g_\pm^{(i)}$ is a  \emph{regularizing} term present at second order only (i.e., $g_\pm^{(1)} = 0$) enforcing the correct asymptotic limits. Indeed, one can check that, were that term not present, the RWZ variables would exhibit more divergent behaviors at second order compared to first-order QNMs~\cite{Gleiser:1995gx,Nicasio_1999,Gleiser_2000,Brizuela:2006ne,Brizuela:2007zza,Brizuela:2009qd,Nakano:2007cj}. Our definition ensures that
\begin{equation} \label{eq:asymptoticQNM}
    \Psi_\pm^{(i)} \underrel{{r \rightarrow \infty}}{\simeq} \mathcal{B}^{(i)}_\pm e^{i \omega (r_*-t)} \; , \; \Psi_\pm^{(i)} \underrel{{r \rightarrow 2M}}{\simeq} \mathcal{C}^{(i)}_\pm e^{-i \omega (r_*+t)} \; ,
\end{equation}
where $\mathcal{B}^{(i)}_\pm$, $\mathcal{C}^{(i)}_\pm$ are constants. At second order, a specific QQNM is generated by the product of two LQNMs $\Psi_{1}^{(1)}$ and $\Psi_{2}^{(1)}$ (for brevity we include the parity of first-order modes in the set of suppressed indices). We thus take
\begin{equation}
    g^{(2)}_\pm = \left(a_{\pm,2}r^2+a_{\pm,1}r\right)\Psi^{(1)}_1\Psi^{(1)}_2
\end{equation}
where $a_{\pm,1},a_{\pm,2}$ are chosen to cancel the divergences of the unregulated master scalars.
Another possibility would have been to work with \textquotedblleft unregularized\textquotedblright{} variables; however, for numerical accuracy we have chosen here to analytically subtract these spurious terms. We have verified that they do not contribute to the asymptotic waveform in TT gauge.
Explicit expressions for $g_\pm^{(2)}$ are given in the companion article~\cite{Bucciotti:2024jrv} {\cb and in the GitHub package~\cite{csvQuadratic}}. Eqs.~\eqref{eq:EinstEqFirstOrder}-\eqref{eq:EinstEqSecondOrder} then translate in the RWZ equations with a source term
\begin{equation}\label{eq:RW}
    \frac{\mathrm{d}^2 \Psi_\pm^{(i)}}{\mathrm{d}r_*^2} + \big( \omega^2 - V_\pm \big) \Psi_\pm^{(i)} = S_\pm^{(i)} \; ,
\end{equation}
where $V_\pm$ are the usual RWZ potentials given in app.~\ref{app:technical} and the source is zero at first order, $S_\pm^{(1)}=0$. Our choice of regulator for $\Psi^{(2)}_\pm$ ensures that the left-hand side of eq.~\eqref{eq:RW} is the same at first and second order. 
On the other hand, the explicit source at second order is regulator-dependent and we get
\begin{align}
    &S^{(2)}_\pm = \mathcal F_1(r) \Psi_1^{(1)} \Psi_2^{(1)} + \mathcal F_2(r)  (\Psi_1^{(1)})' \Psi_2^{(1)} \nonumber \\
    &+ \mathcal F_3(r)   \Psi_1^{(1)} (\Psi_2^{(1)})' + \mathcal F_4(r) (\Psi_1^{(1)})' (\Psi_2^{(1)})' \; , \label{eq:source}
\end{align}
where $'=\mathrm{d}/\mathrm{d}r$, and the four functions $\mathcal F$ depend on $\ell m$, the parities of second and first-order modes as well as $\mathcal{N} = (\ell_1, m_1, n_1, \mathfrak m_1) \times (\ell_2, m_2, n_2, \mathfrak m_2)$. We give their explicit expression in the companion article~\cite{Bucciotti:2024jrv} and in the GitHub package~\cite{csvQuadratic}. We have verified that the asymptotic behavior of LQNMs~\eqref{eq:asymptoticQNM} imposes $S_\pm^{(2)} \propto r^{-2}$ for $r \rightarrow \infty$ and $S_\pm^{(2)} \propto 1-2M/r$ for $r \rightarrow 2M$, which is consistent with eq.~\eqref{eq:RW} at second order once we impose QNM boundary conditions on $\Psi_\pm^{(2)}$. This serves as a confirmation of the consistency of our computations.

\textbf{Checks and results}
\begin{figure}
\includegraphics[width=\columnwidth]{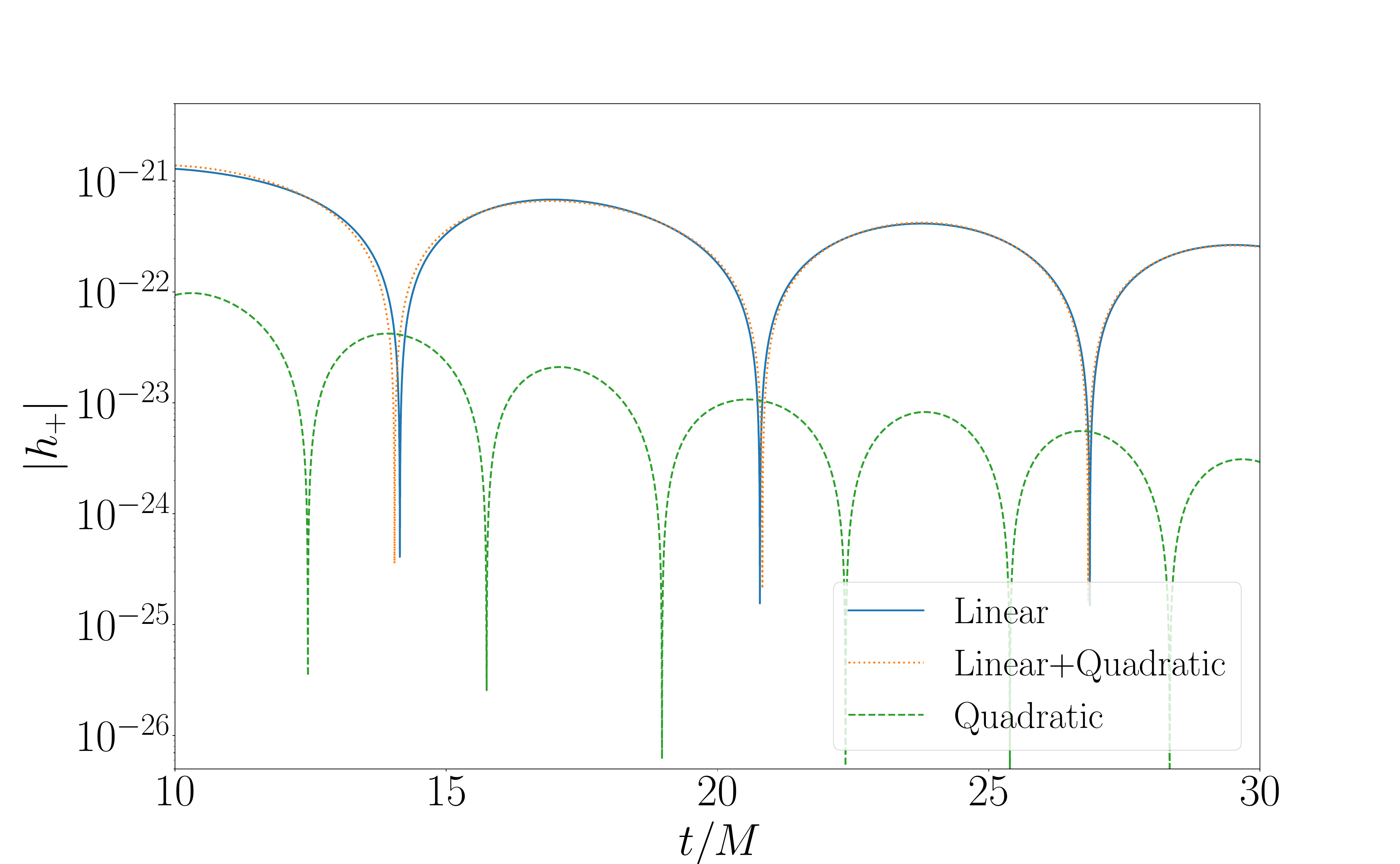}
\caption{\label{fig:waveform} Waveform polarization $|h_+|$ in eq.~\eqref{eq:QNMTT} including only linear modes, only quadratic modes, and both of them for a GW150914-like merger. LQNMs amplitudes are taken from the fits in~\cite{Cheung:2023vki} with progenitor parameters from~\cite{LIGOScientific:2018mvr}; we include the $(220)$, $(221)$, $(210)$, $(330)$ and $(440)$ LQNMs. All quadratic modes amplitudes are added using results from this article. The starting time is chosen to be $t_0=10M$ after the peak of the waveform to ensure QNM regime.}
\end{figure}
Second-order RWZ equations were investigated in several other works, see e.g.~\cite{Nakano:2007cj,Ioka:2007ak,Bucciotti:2023ets,Brizuela:2006ne,Brizuela:2007zza,Brizuela:2009qd,Spiers:2023mor,Ma:2024qcv,Gleiser:1995gx,Nicasio_1999,Gleiser_2000,Riva:2023rcm,Perrone:2023jzq}. However, we cannot directly compare our expression for the source term in~\eqref{eq:source} to their computations since our definitions of master scalars~\eqref{eq:RWvar}-\eqref{eq:Zvar} differ. 
What is unequivocally defined, however, is the asymptotic amplitude of quadratic modes in the GW strain~\eqref{eq:QNMTT}, which is our main result and that we will now provide.

We numerically solve eq.~\eqref{eq:RW} with QNM boundary conditions~\eqref{eq:asymptoticQNM}. We employ the adaptation of the Leaver algorithm to QQNMs described in~\cite{Nakano:2007cj}, which provides the ratio of QQNMs to LQNMs amplitudes at infinity $\mathcal{B}^{(2)}_\pm / \big( \mathcal{B}^{(1)}_1 \mathcal{B}^{(1)}_2 \big)$ with negligible numerical error. Subsequently, we invert eqs.~\eqref{eq:RWvar}-\eqref{eq:Zvar} to reconstruct the gauge-invariant variable $\tilde h$ from the RWZ scalars, which finally provides us the physical amplitude of QNMs in the TT gauge as per eq.~\eqref{eq:QNMTT}. Rotational symmetry imposes that the only dependence of $\mathcal{A}_{\ell m \mathcal{N}}^{(2)}$ on $m, m_1, m_2$ is contained in the 3j-symbol $\begin{pmatrix}
    \ell & \ell_1 & \ell_2 \\ -m & m_1 & m_2
\end{pmatrix}$ which we factor out from our numerical computations. We give in the companion file~\cite{csvQuadratic} the ratios  $\mathcal{R} = \mathcal{A}_{\ell m \mathcal{N}}^{(2)} / \big( \mathcal{A}_{\ell_1 m_1 \mathcal{N}_1}^{(1)} \mathcal{A}_{\ell_2 m_2 \mathcal{N}_2}^{(1)} \big) $ for all $2 \leq \ell, \ell_1, \ell_2 \leq 7$  respecting the Clebsch-Gordan rules $|\ell_1-\ell_2|\leq \ell \leq \ell_1+\ell_2$. We also include the cases where one of the linear amplitudes is an overtone $n=1$ or a mirror mode $\mathfrak m = -$. 

Our results for some of the low-lying modes are displayed in fig.~\ref{fig:ampl}. In particular, we find $\mathcal{R}\simeq0.154e^{-0.068i}$ for $(220+)\times(220+)\rightarrow(44)$ which is the most excited nonlinear mode in NR simulations and BHPT fits~\cite{London_2014,Ma_2022,Mitman:2022qdl,Cheung:2022rbm,Cheung:2023vki,Zhu:2024rej,Redondo-Yuste:2023seq}. This is $\sim 10$\% higher than the semi-analytical calculation on a Kerr background given in~\cite{Ma:2024qcv}. However, our result is in perfect agreement with the fits to NR simulations provided in~\cite{Zhu:2024rej} {\cb at zero spin}\footnote{We refer here to their results based on full NR simulations, which is still $\sim 10$\% lower than their time evolution of second-order perturbation equations. Based on this observation we conjecture that the fits to second-order perturbation theory in~\cite{Zhu:2024rej} contains some source of systematic uncertainties, as was suggested in that article.}. 
Our phase seem also to be in good agreement with~\cite{Zhu:2024rej}.
Another QQNM of interest is $(330+)\times(220+)\rightarrow(55)$ for which we find $\mathcal{R} \simeq 0.417e^{-0.081 i}$, which is a $\sim 7$\% difference in amplitude from the fits provided in~\cite{Cheung:2023vki}, {\cb extrapolated to zero spin}. The phase is however half of the value provided in~\cite{Cheung:2023vki}: this difference may be due to the extrapolation. Finally we find $\mathcal{R} \simeq 0.194 e^{-0.084i}$ for $(220+)\times(210+)\rightarrow(43)$, a QQNM not yet mentioned in the literature.

Let us now describe other peculiarities of our results. First, one has $|\mathcal{R}|<1$ for all of the ratios displayed in fig.~\ref{fig:ampl}; this was not granted since perturbation theory only requires $\mathcal{A}^{(2)}\ll \mathcal{A}^{(1)}$. In addition, it appears that combinations of a regular $\mathfrak m =+$ {\cb and $m>0$} with a mirror $\mathfrak m=-$ {\cb and $m<0$} linear modes give a non-negligible ratio in the $\ell =2$ sector; we find for example $\mathcal{R} \simeq 0.169e^{-0.194i}$ for $(330+)\times(2(\raisebox{.75pt}{-}1)0-) \rightarrow (22)$ and $\mathcal{R} \simeq 0.104e^{-1.58i}$ for $(220+)\times(2(\raisebox{.75pt}{-}1)0-) \rightarrow (21)$.
These QQNMs were not mentioned in previous works using fits to waveforms~\cite{London_2014,Ma_2022,Mitman:2022qdl,Cheung:2022rbm,Cheung:2023vki,Zhu:2024rej,Redondo-Yuste:2023seq}. This may be due to the fact that: a) They have a small or even vanishing real part of the frequency due to the relation $\omega^\mathrm{Q} = \omega_{\ell_1 n_1 +} - (\omega_{\ell_2 n_2 +})^*$, making them potentially more challenging to incorporate into fitting algorithms. b) Their ratio $\mathcal{R}$ is non-negligible mostly for odd $m_1$ or $m_2$, whose parent modes amplitudes are suppressed in equal-mass mergers~\cite{Borhanian:2019kxt,Cheung:2023vki}. 
Lastly, the combination of fundamental $n=0$ and overtone $n=1$ can also present non-negligible ratios $\mathcal{R}$, although these modes are expected to be more challenging to detect due to their shorter decay time. 

{\cb
Additionally, in fig.~\ref{fig:waveform} we illustrate the impact of quadratic modes on a typical ringdown waveform for a GW150914-like merger. From the figure, it is evident that the quadratic oscillations occur at twice the frequency of the linear modes, and the fact that the quadratic amplitude is proportional to the square of the linear one. Notice that, for the purpose of this figure, we have extrapolated our results to the spinning case by simply neglecting the spin dependence of the nonlinear ratio $\mathcal{R}$. This assumption is approximately supported by NR simulations~\cite{Cheung:2023vki,Zhu:2024rej}.  
}

\textbf{Conclusions} In this letter we have provided the first \textit{ab initio} computation of all low multipole QQNMs amplitudes using perturbation theory on a \sch BH background. Our findings should facilitate a straightforward integration of second-order nonlinearities into ringdown models, as QQNMs frequencies and amplitudes are now fully determined by their parent LQNMs. In addition to providing more accurate models for fitting to data or NR, our work enables to test GR by measuring the nonlinear ratio $\mathcal{R}$ in data and comparing it with our prediction, {\cb provided that $\mathcal{R}$ has a weak dependence on spin effects and deviations to reflection symmetry, as suggested by recent works~\cite{Redondo-Yuste:2023seq,Cheung:2023vki,Bourg:2024jme}}. We leave these aspects, as well as the extension of our results to Kerr backgrounds, to future investigations.

\acknowledgments
A. Kuntz acknowledges support from the European Union's H2020 ERC Consolidator Grant ``GRavity from Astrophysical to Microscopic Scales'' (Grant No. GRAMS-815673), the PRIN 2022 grant ``GUVIRP - Gravity tests in the UltraViolet and InfraRed with Pulsar timing'', and the EU Horizon 2020 Research and Innovation Programme under the Marie Sklodowska-Curie Grant Agreement No. 101007855. This project made use of the Black Hole Perturbation Toolkit~\cite{BHPToolkit}. We would like to thank Enrico Barausse and Luca Santoni for discussions.

\textbf{Note added} Our results are compatible with the recent related article~\cite{Bourg:2024jme}, uploaded to the Arxiv just after ours, by setting $C^-_{\ell \texttt{mn}}=0$ in their work, which corresponds to our assumption of equatorial symmetry. While we computed the amplitude of several QQNMs, they concentrated on the most excited QQNM and studied deviations from equatorial symmetry. Their approach could explain the discrepancies we mention when comparing our results to the literature.

\appendix
\section{Technical reference}
\label{app:technical}
\textbf{Spherical harmonics}.
Following the literature, we define tensor spherical harmonics as
\begin{align}
    Y_A^{\ell m} = D_A Y^{\ell m} \; ,\\
    X_A^{\ell m} = -\epsilon_A^{\;\;C} D_C Y^{\ell m} \; ,\\
    Y_{AB}^{\ell m} = \left(D_AD_B+\frac{1}{2}\ell(\ell+1)\Omega_{AB}\right)Y^{\ell m} \; ,\\
    X_{AB}^{\ell m} = -\epsilon_{(A}^{\;\;\;\;C} D_{B)}D_CY^{\ell m} \; ,
\end{align}
where $Y^{\ell m}$ are the standard spherical harmonics, $D_A$ and $\epsilon_{AB}$ are the covariant derivative and the Levi-Civita symbol on the unit sphere respectively.

\textbf{Gauge-invariant perturbations}.
Following~\cite{Spiers:2023mor}, we define
\begin{align}
    \tilde h_{ab}^{(i)} &= h_{ab}^{(i)} + H_{ab}^{(i)} + 2 D_{(a} \zeta^{(i)}{}_{b)} \; , \\
     \tilde h_{a-}^{(i)} &= h_{a-}^{(i)} + H_{a-}^{(i)} + r^2 D_{a} Z_-^{(i)} \; , \\
      \tilde h_{\circ}^{(i)} &= h_{\circ}^{(i)} + H_{\circ}^{(i)} + 2 r f(r)\zeta^{(i)}_r - \ell(\ell+1) r^2 Z_+^{(i)} \; ,
\end{align}
where $Z_\pm^{(i)} = - (h_{\pm}^{(i)} + H_{\pm}^{(i)})/(2r^2)$ and $\zeta_a^{(i)} = - h_{a +}^{(i)} - H_{a+}^{(i)} - r^2 D_a Z_+^{(i)}$. The various $H^{(i)}$ components can be found by decomposing a tensor $H^{(i)}_{\mu \nu}$ in spherical harmonics in the same way as eqs~\eqref{eq:hab}-\eqref{eq:hAB}, where $H^{(1)}_{\mu \nu} = 0$ and $H^{(2)}_{\mu \nu} = \mathcal{L}_\xi \big( h_{\mu \nu}^{(1)} + \mathcal{L}_\xi \bar g_{\mu \nu}/2 \big)$, where $\mathcal{L}$ is the Lie derivative, $\xi_\mu = (\zeta^{(1)}_a, Z^{(1)}_A)$ and $Z_A^{(1)} = Z_+^{(1)} Y_A + Z_-^{(1)} X_A$.

\textbf{RWZ potentials}. They are given by
\begin{align} \label{eq:RWZPot}
    V_- &= f(r) \bigg( \frac{\ell(\ell+1)}{r^2} - \frac{6 M}{r^3} \bigg) \; , \nonumber \\
    V_+ &= f(r) \frac{2 \lambda^2(\lambda+1)r^3 + 6 \lambda^2 M r^2 + 18 \lambda M^2 r + 18 M^3}{r^3(\lambda r + 3 M)^2} \; ,
\end{align}
where $\lambda = (\ell+2)(\ell-1)/2$.
 
\bibliography{bib}

\begin{thebibliography}{68}%
\makeatletter
\providecommand \@ifxundefined [1]{%
 \@ifx{#1\undefined}
}%
\providecommand \@ifnum [1]{%
 \ifnum #1\expandafter \@firstoftwo
 \else \expandafter \@secondoftwo
 \fi
}%
\providecommand \@ifx [1]{%
 \ifx #1\expandafter \@firstoftwo
 \else \expandafter \@secondoftwo
 \fi
}%
\providecommand \natexlab [1]{#1}%
\providecommand \enquote  [1]{``#1''}%
\providecommand \bibnamefont  [1]{#1}%
\providecommand \bibfnamefont [1]{#1}%
\providecommand \citenamefont [1]{#1}%
\providecommand \href@noop [0]{\@secondoftwo}%
\providecommand \href [0]{\begingroup \@sanitize@url \@href}%
\providecommand \@href[1]{\@@startlink{#1}\@@href}%
\providecommand \@@href[1]{\endgroup#1\@@endlink}%
\providecommand \@sanitize@url [0]{\catcode `\\12\catcode `\$12\catcode `\&12\catcode `\#12\catcode `\^12\catcode `\_12\catcode `\%12\relax}%
\providecommand \@@startlink[1]{}%
\providecommand \@@endlink[0]{}%
\providecommand \url  [0]{\begingroup\@sanitize@url \@url }%
\providecommand \@url [1]{\endgroup\@href {#1}{\urlprefix }}%
\providecommand \urlprefix  [0]{URL }%
\providecommand \Eprint [0]{\href }%
\providecommand \doibase [0]{https://doi.org/}%
\providecommand \selectlanguage [0]{\@gobble}%
\providecommand \bibinfo  [0]{\@secondoftwo}%
\providecommand \bibfield  [0]{\@secondoftwo}%
\providecommand \translation [1]{[#1]}%
\providecommand \BibitemOpen [0]{}%
\providecommand \bibitemStop [0]{}%
\providecommand \bibitemNoStop [0]{.\EOS\space}%
\providecommand \EOS [0]{\spacefactor3000\relax}%
\providecommand \BibitemShut  [1]{\csname bibitem#1\endcsname}%
\let\auto@bib@innerbib\@empty
\bibitem [{\citenamefont {Berti}\ \emph {et~al.}(2009)\citenamefont {Berti}, \citenamefont {Cardoso},\ and\ \citenamefont {Starinets}}]{Berti_2009}%
  \BibitemOpen
  \bibfield  {author} {\bibinfo {author} {\bibfnamefont {E.}~\bibnamefont {Berti}}, \bibinfo {author} {\bibfnamefont {V.}~\bibnamefont {Cardoso}},\ and\ \bibinfo {author} {\bibfnamefont {A.~O.}\ \bibnamefont {Starinets}},\ }\bibfield  {title} {\bibinfo {title} {Quasinormal modes of black holes and black branes},\ }\href {https://doi.org/10.1088/0264-9381/26/16/163001} {\bibfield  {journal} {\bibinfo  {journal} {Classical and Quantum Gravity}\ }\textbf {\bibinfo {volume} {26}},\ \bibinfo {pages} {163001} (\bibinfo {year} {2009})}\BibitemShut {NoStop}%
\bibitem [{\citenamefont {Dreyer}\ \emph {et~al.}(2004)\citenamefont {Dreyer}, \citenamefont {Kelly}, \citenamefont {Krishnan}, \citenamefont {Finn}, \citenamefont {Garrison},\ and\ \citenamefont {Lopez-Aleman}}]{Dreyer:2003bv}%
  \BibitemOpen
  \bibfield  {author} {\bibinfo {author} {\bibfnamefont {O.}~\bibnamefont {Dreyer}}, \bibinfo {author} {\bibfnamefont {B.~J.}\ \bibnamefont {Kelly}}, \bibinfo {author} {\bibfnamefont {B.}~\bibnamefont {Krishnan}}, \bibinfo {author} {\bibfnamefont {L.~S.}\ \bibnamefont {Finn}}, \bibinfo {author} {\bibfnamefont {D.}~\bibnamefont {Garrison}},\ and\ \bibinfo {author} {\bibfnamefont {R.}~\bibnamefont {Lopez-Aleman}},\ }\bibfield  {title} {\bibinfo {title} {{Black hole spectroscopy: Testing general relativity through gravitational wave observations}},\ }\href {https://doi.org/10.1088/0264-9381/21/4/003} {\bibfield  {journal} {\bibinfo  {journal} {Class. Quant. Grav.}\ }\textbf {\bibinfo {volume} {21}},\ \bibinfo {pages} {787} (\bibinfo {year} {2004})},\ \Eprint {https://arxiv.org/abs/gr-qc/0309007} {arXiv:gr-qc/0309007} \BibitemShut {NoStop}%
\bibitem [{\citenamefont {Berti}\ \emph {et~al.}(2006)\citenamefont {Berti}, \citenamefont {Cardoso},\ and\ \citenamefont {Will}}]{Berti_2006}%
  \BibitemOpen
  \bibfield  {author} {\bibinfo {author} {\bibfnamefont {E.}~\bibnamefont {Berti}}, \bibinfo {author} {\bibfnamefont {V.}~\bibnamefont {Cardoso}},\ and\ \bibinfo {author} {\bibfnamefont {C.~M.}\ \bibnamefont {Will}},\ }\bibfield  {title} {\bibinfo {title} {Gravitational-wave spectroscopy of massive black holes with the space interferometer lisa},\ }\bibfield  {journal} {\bibinfo  {journal} {Physical Review D}\ }\textbf {\bibinfo {volume} {73}},\ \href {https://doi.org/10.1103/physrevd.73.064030} {10.1103/physrevd.73.064030} (\bibinfo {year} {2006})\BibitemShut {NoStop}%
\bibitem [{\citenamefont {Berti}\ \emph {et~al.}(2016)\citenamefont {Berti}, \citenamefont {Sesana}, \citenamefont {Barausse}, \citenamefont {Cardoso},\ and\ \citenamefont {Belczynski}}]{Berti_2016}%
  \BibitemOpen
  \bibfield  {author} {\bibinfo {author} {\bibfnamefont {E.}~\bibnamefont {Berti}}, \bibinfo {author} {\bibfnamefont {A.}~\bibnamefont {Sesana}}, \bibinfo {author} {\bibfnamefont {E.}~\bibnamefont {Barausse}}, \bibinfo {author} {\bibfnamefont {V.}~\bibnamefont {Cardoso}},\ and\ \bibinfo {author} {\bibfnamefont {K.}~\bibnamefont {Belczynski}},\ }\bibfield  {title} {\bibinfo {title} {Spectroscopy of kerr black holes with earth- and space-based interferometers},\ }\bibfield  {journal} {\bibinfo  {journal} {Physical Review Letters}\ }\textbf {\bibinfo {volume} {117}},\ \href {https://doi.org/10.1103/physrevlett.117.101102} {10.1103/physrevlett.117.101102} (\bibinfo {year} {2016})\BibitemShut {NoStop}%
\bibitem [{\citenamefont {Bhagwat}\ \emph {et~al.}(2016)\citenamefont {Bhagwat}, \citenamefont {Brown},\ and\ \citenamefont {Ballmer}}]{Bhagwat:2016ntk}%
  \BibitemOpen
  \bibfield  {author} {\bibinfo {author} {\bibfnamefont {S.}~\bibnamefont {Bhagwat}}, \bibinfo {author} {\bibfnamefont {D.~A.}\ \bibnamefont {Brown}},\ and\ \bibinfo {author} {\bibfnamefont {S.~W.}\ \bibnamefont {Ballmer}},\ }\bibfield  {title} {\bibinfo {title} {{Spectroscopic analysis of stellar mass black-hole mergers in our local universe with ground-based gravitational wave detectors}},\ }\href {https://doi.org/10.1103/PhysRevD.94.084024} {\bibfield  {journal} {\bibinfo  {journal} {Phys. Rev. D}\ }\textbf {\bibinfo {volume} {94}},\ \bibinfo {pages} {084024} (\bibinfo {year} {2016})},\ \bibinfo {note} {[Erratum: Phys.Rev.D 95, 069906 (2017)]},\ \Eprint {https://arxiv.org/abs/1607.07845} {arXiv:1607.07845 [gr-qc]} \BibitemShut {NoStop}%
\bibitem [{\citenamefont {Yang}\ \emph {et~al.}(2017)\citenamefont {Yang}, \citenamefont {Yagi}, \citenamefont {Blackman}, \citenamefont {Lehner}, \citenamefont {Paschalidis}, \citenamefont {Pretorius},\ and\ \citenamefont {Yunes}}]{Yang:2017zxs}%
  \BibitemOpen
  \bibfield  {author} {\bibinfo {author} {\bibfnamefont {H.}~\bibnamefont {Yang}}, \bibinfo {author} {\bibfnamefont {K.}~\bibnamefont {Yagi}}, \bibinfo {author} {\bibfnamefont {J.}~\bibnamefont {Blackman}}, \bibinfo {author} {\bibfnamefont {L.}~\bibnamefont {Lehner}}, \bibinfo {author} {\bibfnamefont {V.}~\bibnamefont {Paschalidis}}, \bibinfo {author} {\bibfnamefont {F.}~\bibnamefont {Pretorius}},\ and\ \bibinfo {author} {\bibfnamefont {N.}~\bibnamefont {Yunes}},\ }\bibfield  {title} {\bibinfo {title} {{Black hole spectroscopy with coherent mode stacking}},\ }\href {https://doi.org/10.1103/PhysRevLett.118.161101} {\bibfield  {journal} {\bibinfo  {journal} {Phys. Rev. Lett.}\ }\textbf {\bibinfo {volume} {118}},\ \bibinfo {pages} {161101} (\bibinfo {year} {2017})},\ \Eprint {https://arxiv.org/abs/1701.05808} {arXiv:1701.05808 [gr-qc]} \BibitemShut {NoStop}%
\bibitem [{\citenamefont {Baibhav}\ \emph {et~al.}(2018)\citenamefont {Baibhav}, \citenamefont {Berti}, \citenamefont {Cardoso},\ and\ \citenamefont {Khanna}}]{Baibhav:2017jhs}%
  \BibitemOpen
  \bibfield  {author} {\bibinfo {author} {\bibfnamefont {V.}~\bibnamefont {Baibhav}}, \bibinfo {author} {\bibfnamefont {E.}~\bibnamefont {Berti}}, \bibinfo {author} {\bibfnamefont {V.}~\bibnamefont {Cardoso}},\ and\ \bibinfo {author} {\bibfnamefont {G.}~\bibnamefont {Khanna}},\ }\bibfield  {title} {\bibinfo {title} {{Black Hole Spectroscopy: Systematic Errors and Ringdown Energy Estimates}},\ }\href {https://doi.org/10.1103/PhysRevD.97.044048} {\bibfield  {journal} {\bibinfo  {journal} {Phys. Rev. D}\ }\textbf {\bibinfo {volume} {97}},\ \bibinfo {pages} {044048} (\bibinfo {year} {2018})},\ \Eprint {https://arxiv.org/abs/1710.02156} {arXiv:1710.02156 [gr-qc]} \BibitemShut {NoStop}%
\bibitem [{\citenamefont {Brito}\ \emph {et~al.}(2018)\citenamefont {Brito}, \citenamefont {Buonanno},\ and\ \citenamefont {Raymond}}]{Brito:2018rfr}%
  \BibitemOpen
  \bibfield  {author} {\bibinfo {author} {\bibfnamefont {R.}~\bibnamefont {Brito}}, \bibinfo {author} {\bibfnamefont {A.}~\bibnamefont {Buonanno}},\ and\ \bibinfo {author} {\bibfnamefont {V.}~\bibnamefont {Raymond}},\ }\bibfield  {title} {\bibinfo {title} {{Black-hole Spectroscopy by Making Full Use of Gravitational-Wave Modeling}},\ }\href {https://doi.org/10.1103/PhysRevD.98.084038} {\bibfield  {journal} {\bibinfo  {journal} {Phys. Rev. D}\ }\textbf {\bibinfo {volume} {98}},\ \bibinfo {pages} {084038} (\bibinfo {year} {2018})},\ \Eprint {https://arxiv.org/abs/1805.00293} {arXiv:1805.00293 [gr-qc]} \BibitemShut {NoStop}%
\bibitem [{\citenamefont {Baibhav}\ and\ \citenamefont {Berti}(2019)}]{Baibhav:2018rfk}%
  \BibitemOpen
  \bibfield  {author} {\bibinfo {author} {\bibfnamefont {V.}~\bibnamefont {Baibhav}}\ and\ \bibinfo {author} {\bibfnamefont {E.}~\bibnamefont {Berti}},\ }\bibfield  {title} {\bibinfo {title} {{Multimode black hole spectroscopy}},\ }\href {https://doi.org/10.1103/PhysRevD.99.024005} {\bibfield  {journal} {\bibinfo  {journal} {Phys. Rev. D}\ }\textbf {\bibinfo {volume} {99}},\ \bibinfo {pages} {024005} (\bibinfo {year} {2019})},\ \Eprint {https://arxiv.org/abs/1809.03500} {arXiv:1809.03500 [gr-qc]} \BibitemShut {NoStop}%
\bibitem [{\citenamefont {Bhagwat}\ \emph {et~al.}(2020)\citenamefont {Bhagwat}, \citenamefont {Forteza}, \citenamefont {Pani},\ and\ \citenamefont {Ferrari}}]{Bhagwat:2019dtm}%
  \BibitemOpen
  \bibfield  {author} {\bibinfo {author} {\bibfnamefont {S.}~\bibnamefont {Bhagwat}}, \bibinfo {author} {\bibfnamefont {X.~J.}\ \bibnamefont {Forteza}}, \bibinfo {author} {\bibfnamefont {P.}~\bibnamefont {Pani}},\ and\ \bibinfo {author} {\bibfnamefont {V.}~\bibnamefont {Ferrari}},\ }\bibfield  {title} {\bibinfo {title} {{Ringdown overtones, black hole spectroscopy, and no-hair theorem tests}},\ }\href {https://doi.org/10.1103/PhysRevD.101.044033} {\bibfield  {journal} {\bibinfo  {journal} {Phys. Rev. D}\ }\textbf {\bibinfo {volume} {101}},\ \bibinfo {pages} {044033} (\bibinfo {year} {2020})},\ \Eprint {https://arxiv.org/abs/1910.08708} {arXiv:1910.08708 [gr-qc]} \BibitemShut {NoStop}%
\bibitem [{\citenamefont {Maselli}\ \emph {et~al.}(2020)\citenamefont {Maselli}, \citenamefont {Pani}, \citenamefont {Gualtieri},\ and\ \citenamefont {Berti}}]{Maselli_2020}%
  \BibitemOpen
  \bibfield  {author} {\bibinfo {author} {\bibfnamefont {A.}~\bibnamefont {Maselli}}, \bibinfo {author} {\bibfnamefont {P.}~\bibnamefont {Pani}}, \bibinfo {author} {\bibfnamefont {L.}~\bibnamefont {Gualtieri}},\ and\ \bibinfo {author} {\bibfnamefont {E.}~\bibnamefont {Berti}},\ }\bibfield  {title} {\bibinfo {title} {Parametrized ringdown spin expansion coefficients: A data-analysis framework for black-hole spectroscopy with multiple events},\ }\bibfield  {journal} {\bibinfo  {journal} {Physical Review D}\ }\textbf {\bibinfo {volume} {101}},\ \href {https://doi.org/10.1103/physrevd.101.024043} {10.1103/physrevd.101.024043} (\bibinfo {year} {2020})\BibitemShut {NoStop}%
\bibitem [{\citenamefont {Calder\'on~Bustillo}\ \emph {et~al.}(2021)\citenamefont {Calder\'on~Bustillo}, \citenamefont {Lasky},\ and\ \citenamefont {Thrane}}]{CalderonBustillo:2020rmh}%
  \BibitemOpen
  \bibfield  {author} {\bibinfo {author} {\bibfnamefont {J.}~\bibnamefont {Calder\'on~Bustillo}}, \bibinfo {author} {\bibfnamefont {P.~D.}\ \bibnamefont {Lasky}},\ and\ \bibinfo {author} {\bibfnamefont {E.}~\bibnamefont {Thrane}},\ }\bibfield  {title} {\bibinfo {title} {{Black-hole spectroscopy, the no-hair theorem, and GW150914: Kerr versus Occam}},\ }\href {https://doi.org/10.1103/PhysRevD.103.024041} {\bibfield  {journal} {\bibinfo  {journal} {Phys. Rev. D}\ }\textbf {\bibinfo {volume} {103}},\ \bibinfo {pages} {024041} (\bibinfo {year} {2021})},\ \Eprint {https://arxiv.org/abs/2010.01857} {arXiv:2010.01857 [gr-qc]} \BibitemShut {NoStop}%
\bibitem [{\citenamefont {Capano}\ and\ \citenamefont {Nitz}(2020)}]{Capano:2020dix}%
  \BibitemOpen
  \bibfield  {author} {\bibinfo {author} {\bibfnamefont {C.~D.}\ \bibnamefont {Capano}}\ and\ \bibinfo {author} {\bibfnamefont {A.~H.}\ \bibnamefont {Nitz}},\ }\bibfield  {title} {\bibinfo {title} {{Binary black hole spectroscopy: a no-hair test of GW190814 and GW190412}},\ }\href {https://doi.org/10.1103/PhysRevD.102.124070} {\bibfield  {journal} {\bibinfo  {journal} {Phys. Rev. D}\ }\textbf {\bibinfo {volume} {102}},\ \bibinfo {pages} {124070} (\bibinfo {year} {2020})},\ \Eprint {https://arxiv.org/abs/2008.02248} {arXiv:2008.02248 [gr-qc]} \BibitemShut {NoStop}%
\bibitem [{\citenamefont {Jim\'enez~Forteza}\ \emph {et~al.}(2020)\citenamefont {Jim\'enez~Forteza}, \citenamefont {Bhagwat}, \citenamefont {Pani},\ and\ \citenamefont {Ferrari}}]{JimenezForteza:2020cve}%
  \BibitemOpen
  \bibfield  {author} {\bibinfo {author} {\bibfnamefont {X.}~\bibnamefont {Jim\'enez~Forteza}}, \bibinfo {author} {\bibfnamefont {S.}~\bibnamefont {Bhagwat}}, \bibinfo {author} {\bibfnamefont {P.}~\bibnamefont {Pani}},\ and\ \bibinfo {author} {\bibfnamefont {V.}~\bibnamefont {Ferrari}},\ }\bibfield  {title} {\bibinfo {title} {{Spectroscopy of binary black hole ringdown using overtones and angular modes}},\ }\href {https://doi.org/10.1103/PhysRevD.102.044053} {\bibfield  {journal} {\bibinfo  {journal} {Phys. Rev. D}\ }\textbf {\bibinfo {volume} {102}},\ \bibinfo {pages} {044053} (\bibinfo {year} {2020})},\ \Eprint {https://arxiv.org/abs/2005.03260} {arXiv:2005.03260 [gr-qc]} \BibitemShut {NoStop}%
\bibitem [{\citenamefont {Ota}\ and\ \citenamefont {Chirenti}(2021)}]{Ota:2021ypb}%
  \BibitemOpen
  \bibfield  {author} {\bibinfo {author} {\bibfnamefont {I.}~\bibnamefont {Ota}}\ and\ \bibinfo {author} {\bibfnamefont {C.}~\bibnamefont {Chirenti}},\ }\bibfield  {title} {\bibinfo {title} {{Black hole spectroscopy horizons for current and future gravitational wave detectors}},\ }\href@noop {} {\  (\bibinfo {year} {2021})},\ \Eprint {https://arxiv.org/abs/2108.01774} {arXiv:2108.01774 [gr-qc]} \BibitemShut {NoStop}%
\bibitem [{\citenamefont {Maselli}\ \emph {et~al.}(2023)\citenamefont {Maselli}, \citenamefont {Yi}, \citenamefont {Pierini}, \citenamefont {Vellucci}, \citenamefont {Reali}, \citenamefont {Gualtieri},\ and\ \citenamefont {Berti}}]{Maselli:2023khq}%
  \BibitemOpen
  \bibfield  {author} {\bibinfo {author} {\bibfnamefont {A.}~\bibnamefont {Maselli}}, \bibinfo {author} {\bibfnamefont {S.}~\bibnamefont {Yi}}, \bibinfo {author} {\bibfnamefont {L.}~\bibnamefont {Pierini}}, \bibinfo {author} {\bibfnamefont {V.}~\bibnamefont {Vellucci}}, \bibinfo {author} {\bibfnamefont {L.}~\bibnamefont {Reali}}, \bibinfo {author} {\bibfnamefont {L.}~\bibnamefont {Gualtieri}},\ and\ \bibinfo {author} {\bibfnamefont {E.}~\bibnamefont {Berti}},\ }\bibfield  {title} {\bibinfo {title} {{Black hole spectroscopy beyond Kerr: agnostic and theory-based tests with next-generation interferometers}},\ }\href@noop {} {\  (\bibinfo {year} {2023})},\ \Eprint {https://arxiv.org/abs/2311.14803} {arXiv:2311.14803 [gr-qc]} \BibitemShut {NoStop}%
\bibitem [{\citenamefont {Baibhav}\ \emph {et~al.}(2023)\citenamefont {Baibhav}, \citenamefont {Cheung}, \citenamefont {Berti}, \citenamefont {Cardoso}, \citenamefont {Carullo}, \citenamefont {Cotesta}, \citenamefont {Del~Pozzo},\ and\ \citenamefont {Duque}}]{Baibhav:2023clw}%
  \BibitemOpen
  \bibfield  {author} {\bibinfo {author} {\bibfnamefont {V.}~\bibnamefont {Baibhav}}, \bibinfo {author} {\bibfnamefont {M.~H.-Y.}\ \bibnamefont {Cheung}}, \bibinfo {author} {\bibfnamefont {E.}~\bibnamefont {Berti}}, \bibinfo {author} {\bibfnamefont {V.}~\bibnamefont {Cardoso}}, \bibinfo {author} {\bibfnamefont {G.}~\bibnamefont {Carullo}}, \bibinfo {author} {\bibfnamefont {R.}~\bibnamefont {Cotesta}}, \bibinfo {author} {\bibfnamefont {W.}~\bibnamefont {Del~Pozzo}},\ and\ \bibinfo {author} {\bibfnamefont {F.}~\bibnamefont {Duque}},\ }\bibfield  {title} {\bibinfo {title} {{Agnostic black hole spectroscopy: Quasinormal mode content of numerical relativity waveforms and limits of validity of linear perturbation theory}},\ }\href {https://doi.org/10.1103/PhysRevD.108.104020} {\bibfield  {journal} {\bibinfo  {journal} {Phys. Rev. D}\ }\textbf {\bibinfo {volume} {108}},\ \bibinfo {pages} {104020} (\bibinfo {year} {2023})},\ \Eprint {https://arxiv.org/abs/2302.03050} {arXiv:2302.03050 [gr-qc]} \BibitemShut {NoStop}%
\bibitem [{\citenamefont {{Cardoso}}\ \emph {et~al.}(2019)\citenamefont {{Cardoso}}, \citenamefont {{Kimura}}, \citenamefont {{Maselli}}, \citenamefont {{Berti}}, \citenamefont {{Macedo}},\ and\ \citenamefont {{McManus}}}]{2019PhRvD..99j4077C}%
  \BibitemOpen
  \bibfield  {author} {\bibinfo {author} {\bibfnamefont {V.}~\bibnamefont {{Cardoso}}}, \bibinfo {author} {\bibfnamefont {M.}~\bibnamefont {{Kimura}}}, \bibinfo {author} {\bibfnamefont {A.}~\bibnamefont {{Maselli}}}, \bibinfo {author} {\bibfnamefont {E.}~\bibnamefont {{Berti}}}, \bibinfo {author} {\bibfnamefont {C.~F.~B.}\ \bibnamefont {{Macedo}}},\ and\ \bibinfo {author} {\bibfnamefont {R.}~\bibnamefont {{McManus}}},\ }\bibfield  {title} {\bibinfo {title} {{Parametrized black hole quasinormal ringdown: Decoupled equations for nonrotating black holes}},\ }\href {https://doi.org/10.1103/PhysRevD.99.104077} {\bibfield  {journal} {\bibinfo  {journal} {Physical Review D}\ }\textbf {\bibinfo {volume} {99}},\ \bibinfo {eid} {104077} (\bibinfo {year} {2019})},\ \Eprint {https://arxiv.org/abs/1901.01265} {arXiv:1901.01265 [gr-qc]} \BibitemShut {NoStop}%
\bibitem [{\citenamefont {{McManus}}\ \emph {et~al.}(2019)\citenamefont {{McManus}}, \citenamefont {{Berti}}, \citenamefont {{Macedo}}, \citenamefont {{Kimura}}, \citenamefont {{Maselli}},\ and\ \citenamefont {{Cardoso}}}]{2019PhRvD.100d4061M}%
  \BibitemOpen
  \bibfield  {author} {\bibinfo {author} {\bibfnamefont {R.}~\bibnamefont {{McManus}}}, \bibinfo {author} {\bibfnamefont {E.}~\bibnamefont {{Berti}}}, \bibinfo {author} {\bibfnamefont {C.~F.}\ \bibnamefont {{Macedo}}}, \bibinfo {author} {\bibfnamefont {M.}~\bibnamefont {{Kimura}}}, \bibinfo {author} {\bibfnamefont {A.}~\bibnamefont {{Maselli}}},\ and\ \bibinfo {author} {\bibfnamefont {V.}~\bibnamefont {{Cardoso}}},\ }\bibfield  {title} {\bibinfo {title} {{Parametrized black hole quasinormal ringdown. II. Coupled equations and quadratic corrections for nonrotating black holes}},\ }\href {https://doi.org/10.1103/PhysRevD.100.044061} {\bibfield  {journal} {\bibinfo  {journal} {Physical Review D}\ }\textbf {\bibinfo {volume} {100}},\ \bibinfo {eid} {044061} (\bibinfo {year} {2019})},\ \Eprint {https://arxiv.org/abs/1906.05155} {arXiv:1906.05155 [gr-qc]} \BibitemShut {NoStop}%
\bibitem [{\citenamefont {Franchini}\ and\ \citenamefont {V\"olkel}(2023{\natexlab{a}})}]{Franchini:2022axs}%
  \BibitemOpen
  \bibfield  {author} {\bibinfo {author} {\bibfnamefont {N.}~\bibnamefont {Franchini}}\ and\ \bibinfo {author} {\bibfnamefont {S.~H.}\ \bibnamefont {V\"olkel}},\ }\bibfield  {title} {\bibinfo {title} {{Parametrized quasinormal mode framework for non-Schwarzschild metrics}},\ }\href {https://doi.org/10.1103/PhysRevD.107.124063} {\bibfield  {journal} {\bibinfo  {journal} {Phys. Rev. D}\ }\textbf {\bibinfo {volume} {107}},\ \bibinfo {pages} {124063} (\bibinfo {year} {2023}{\natexlab{a}})},\ \Eprint {https://arxiv.org/abs/2210.14020} {arXiv:2210.14020 [gr-qc]} \BibitemShut {NoStop}%
\bibitem [{\citenamefont {Silva}\ \emph {et~al.}(2024)\citenamefont {Silva}, \citenamefont {Tambalo}, \citenamefont {Glampedakis}, \citenamefont {Yagi},\ and\ \citenamefont {Steinhoff}}]{Silva:2024ffz}%
  \BibitemOpen
  \bibfield  {author} {\bibinfo {author} {\bibfnamefont {H.~O.}\ \bibnamefont {Silva}}, \bibinfo {author} {\bibfnamefont {G.}~\bibnamefont {Tambalo}}, \bibinfo {author} {\bibfnamefont {K.}~\bibnamefont {Glampedakis}}, \bibinfo {author} {\bibfnamefont {K.}~\bibnamefont {Yagi}},\ and\ \bibinfo {author} {\bibfnamefont {J.}~\bibnamefont {Steinhoff}},\ }\bibfield  {title} {\bibinfo {title} {{Quasinormal modes and their excitation beyond general relativity}},\ }\href@noop {} {\  (\bibinfo {year} {2024})},\ \Eprint {https://arxiv.org/abs/2404.11110} {arXiv:2404.11110 [gr-qc]} \BibitemShut {NoStop}%
\bibitem [{\citenamefont {Ghosh}\ \emph {et~al.}(2021)\citenamefont {Ghosh}, \citenamefont {Brito},\ and\ \citenamefont {Buonanno}}]{Ghosh:2021mrv}%
  \BibitemOpen
  \bibfield  {author} {\bibinfo {author} {\bibfnamefont {A.}~\bibnamefont {Ghosh}}, \bibinfo {author} {\bibfnamefont {R.}~\bibnamefont {Brito}},\ and\ \bibinfo {author} {\bibfnamefont {A.}~\bibnamefont {Buonanno}},\ }\bibfield  {title} {\bibinfo {title} {{Constraints on quasinormal-mode frequencies with LIGO-Virgo binary\textendash{}black-hole observations}},\ }\href {https://doi.org/10.1103/PhysRevD.103.124041} {\bibfield  {journal} {\bibinfo  {journal} {Phys. Rev. D}\ }\textbf {\bibinfo {volume} {103}},\ \bibinfo {pages} {124041} (\bibinfo {year} {2021})},\ \Eprint {https://arxiv.org/abs/2104.01906} {arXiv:2104.01906 [gr-qc]} \BibitemShut {NoStop}%
\bibitem [{\citenamefont {Abbott}\ \emph {et~al.}(2016)\citenamefont {Abbott} \emph {et~al.}}]{LIGOScientific:2016lio}%
  \BibitemOpen
  \bibfield  {author} {\bibinfo {author} {\bibfnamefont {B.~P.}\ \bibnamefont {Abbott}} \emph {et~al.} (\bibinfo {collaboration} {LIGO Scientific, Virgo}),\ }\bibfield  {title} {\bibinfo {title} {{Tests of general relativity with GW150914}},\ }\href {https://doi.org/10.1103/PhysRevLett.116.221101} {\bibfield  {journal} {\bibinfo  {journal} {Phys. Rev. Lett.}\ }\textbf {\bibinfo {volume} {116}},\ \bibinfo {pages} {221101} (\bibinfo {year} {2016})},\ \bibinfo {note} {[Erratum: Phys.Rev.Lett. 121, 129902 (2018)]},\ \Eprint {https://arxiv.org/abs/1602.03841} {arXiv:1602.03841 [gr-qc]} \BibitemShut {NoStop}%
\bibitem [{\citenamefont {Abbott}\ \emph {et~al.}(2021)\citenamefont {Abbott} \emph {et~al.}}]{LIGOScientific:2020tif}%
  \BibitemOpen
  \bibfield  {author} {\bibinfo {author} {\bibfnamefont {R.}~\bibnamefont {Abbott}} \emph {et~al.} (\bibinfo {collaboration} {LIGO Scientific, Virgo}),\ }\bibfield  {title} {\bibinfo {title} {{Tests of general relativity with binary black holes from the second LIGO-Virgo gravitational-wave transient catalog}},\ }\href {https://doi.org/10.1103/PhysRevD.103.122002} {\bibfield  {journal} {\bibinfo  {journal} {Phys. Rev. D}\ }\textbf {\bibinfo {volume} {103}},\ \bibinfo {pages} {122002} (\bibinfo {year} {2021})},\ \Eprint {https://arxiv.org/abs/2010.14529} {arXiv:2010.14529 [gr-qc]} \BibitemShut {NoStop}%
\bibitem [{\citenamefont {Franchini}\ and\ \citenamefont {V\"olkel}(2023{\natexlab{b}})}]{Franchini:2023eda}%
  \BibitemOpen
  \bibfield  {author} {\bibinfo {author} {\bibfnamefont {N.}~\bibnamefont {Franchini}}\ and\ \bibinfo {author} {\bibfnamefont {S.~H.}\ \bibnamefont {V\"olkel}},\ }\bibfield  {title} {\bibinfo {title} {{Testing General Relativity with Black Hole Quasi-Normal Modes}},\ }\href@noop {} {\  (\bibinfo {year} {2023}{\natexlab{b}})},\ \Eprint {https://arxiv.org/abs/2305.01696} {arXiv:2305.01696 [gr-qc]} \BibitemShut {NoStop}%
\bibitem [{\citenamefont {Toubiana}\ \emph {et~al.}(2024)\citenamefont {Toubiana}, \citenamefont {Pompili}, \citenamefont {Buonanno}, \citenamefont {Gair},\ and\ \citenamefont {Katz}}]{Toubiana:2023cwr}%
  \BibitemOpen
  \bibfield  {author} {\bibinfo {author} {\bibfnamefont {A.}~\bibnamefont {Toubiana}}, \bibinfo {author} {\bibfnamefont {L.}~\bibnamefont {Pompili}}, \bibinfo {author} {\bibfnamefont {A.}~\bibnamefont {Buonanno}}, \bibinfo {author} {\bibfnamefont {J.~R.}\ \bibnamefont {Gair}},\ and\ \bibinfo {author} {\bibfnamefont {M.~L.}\ \bibnamefont {Katz}},\ }\bibfield  {title} {\bibinfo {title} {{Measuring source properties and quasinormal mode frequencies of heavy massive black-hole binaries with LISA}},\ }\href {https://doi.org/10.1103/PhysRevD.109.104019} {\bibfield  {journal} {\bibinfo  {journal} {Phys. Rev. D}\ }\textbf {\bibinfo {volume} {109}},\ \bibinfo {pages} {104019} (\bibinfo {year} {2024})},\ \Eprint {https://arxiv.org/abs/2307.15086} {arXiv:2307.15086 [gr-qc]} \BibitemShut {NoStop}%
\bibitem [{\citenamefont {Hirano}\ \emph {et~al.}(2024)\citenamefont {Hirano}, \citenamefont {Kimura}, \citenamefont {Yamaguchi},\ and\ \citenamefont {Zhang}}]{Hirano:2024fgp}%
  \BibitemOpen
  \bibfield  {author} {\bibinfo {author} {\bibfnamefont {S.}~\bibnamefont {Hirano}}, \bibinfo {author} {\bibfnamefont {M.}~\bibnamefont {Kimura}}, \bibinfo {author} {\bibfnamefont {M.}~\bibnamefont {Yamaguchi}},\ and\ \bibinfo {author} {\bibfnamefont {J.}~\bibnamefont {Zhang}},\ }\bibfield  {title} {\bibinfo {title} {{Parametrized Black Hole Quasinormal Ringdown Formalism for Higher Overtones}},\ }\href@noop {} {\  (\bibinfo {year} {2024})},\ \Eprint {https://arxiv.org/abs/2404.09672} {arXiv:2404.09672 [gr-qc]} \BibitemShut {NoStop}%
\bibitem [{\citenamefont {Buonanno}\ \emph {et~al.}(2007)\citenamefont {Buonanno}, \citenamefont {Cook},\ and\ \citenamefont {Pretorius}}]{Buonanno:2006ui}%
  \BibitemOpen
  \bibfield  {author} {\bibinfo {author} {\bibfnamefont {A.}~\bibnamefont {Buonanno}}, \bibinfo {author} {\bibfnamefont {G.~B.}\ \bibnamefont {Cook}},\ and\ \bibinfo {author} {\bibfnamefont {F.}~\bibnamefont {Pretorius}},\ }\bibfield  {title} {\bibinfo {title} {{Inspiral, merger and ring-down of equal-mass black-hole binaries}},\ }\href {https://doi.org/10.1103/PhysRevD.75.124018} {\bibfield  {journal} {\bibinfo  {journal} {Phys. Rev. D}\ }\textbf {\bibinfo {volume} {75}},\ \bibinfo {pages} {124018} (\bibinfo {year} {2007})},\ \Eprint {https://arxiv.org/abs/gr-qc/0610122} {arXiv:gr-qc/0610122} \BibitemShut {NoStop}%
\bibitem [{\citenamefont {Berti}\ \emph {et~al.}(2007{\natexlab{a}})\citenamefont {Berti}, \citenamefont {Cardoso}, \citenamefont {Gonzalez}, \citenamefont {Sperhake}, \citenamefont {Hannam}, \citenamefont {Husa},\ and\ \citenamefont {Bruegmann}}]{Berti:2007fi}%
  \BibitemOpen
  \bibfield  {author} {\bibinfo {author} {\bibfnamefont {E.}~\bibnamefont {Berti}}, \bibinfo {author} {\bibfnamefont {V.}~\bibnamefont {Cardoso}}, \bibinfo {author} {\bibfnamefont {J.~A.}\ \bibnamefont {Gonzalez}}, \bibinfo {author} {\bibfnamefont {U.}~\bibnamefont {Sperhake}}, \bibinfo {author} {\bibfnamefont {M.}~\bibnamefont {Hannam}}, \bibinfo {author} {\bibfnamefont {S.}~\bibnamefont {Husa}},\ and\ \bibinfo {author} {\bibfnamefont {B.}~\bibnamefont {Bruegmann}},\ }\bibfield  {title} {\bibinfo {title} {{Inspiral, merger and ringdown of unequal mass black hole binaries: A Multipolar analysis}},\ }\href {https://doi.org/10.1103/PhysRevD.76.064034} {\bibfield  {journal} {\bibinfo  {journal} {Phys. Rev. D}\ }\textbf {\bibinfo {volume} {76}},\ \bibinfo {pages} {064034} (\bibinfo {year} {2007}{\natexlab{a}})},\ \Eprint {https://arxiv.org/abs/gr-qc/0703053} {arXiv:gr-qc/0703053} \BibitemShut {NoStop}%
\bibitem [{\citenamefont {Berti}\ \emph {et~al.}(2007{\natexlab{b}})\citenamefont {Berti}, \citenamefont {Cardoso}, \citenamefont {Cardoso},\ and\ \citenamefont {Cavaglia}}]{Berti:2007zu}%
  \BibitemOpen
  \bibfield  {author} {\bibinfo {author} {\bibfnamefont {E.}~\bibnamefont {Berti}}, \bibinfo {author} {\bibfnamefont {J.}~\bibnamefont {Cardoso}}, \bibinfo {author} {\bibfnamefont {V.}~\bibnamefont {Cardoso}},\ and\ \bibinfo {author} {\bibfnamefont {M.}~\bibnamefont {Cavaglia}},\ }\bibfield  {title} {\bibinfo {title} {{Matched-filtering and parameter estimation of ringdown waveforms}},\ }\href {https://doi.org/10.1103/PhysRevD.76.104044} {\bibfield  {journal} {\bibinfo  {journal} {Phys. Rev. D}\ }\textbf {\bibinfo {volume} {76}},\ \bibinfo {pages} {104044} (\bibinfo {year} {2007}{\natexlab{b}})},\ \Eprint {https://arxiv.org/abs/0707.1202} {arXiv:0707.1202 [gr-qc]} \BibitemShut {NoStop}%
\bibitem [{\citenamefont {Giesler}\ \emph {et~al.}(2019)\citenamefont {Giesler}, \citenamefont {Isi}, \citenamefont {Scheel},\ and\ \citenamefont {Teukolsky}}]{Giesler:2019uxc}%
  \BibitemOpen
  \bibfield  {author} {\bibinfo {author} {\bibfnamefont {M.}~\bibnamefont {Giesler}}, \bibinfo {author} {\bibfnamefont {M.}~\bibnamefont {Isi}}, \bibinfo {author} {\bibfnamefont {M.~A.}\ \bibnamefont {Scheel}},\ and\ \bibinfo {author} {\bibfnamefont {S.}~\bibnamefont {Teukolsky}},\ }\bibfield  {title} {\bibinfo {title} {{Black Hole Ringdown: The Importance of Overtones}},\ }\href {https://doi.org/10.1103/PhysRevX.9.041060} {\bibfield  {journal} {\bibinfo  {journal} {Phys. Rev. X}\ }\textbf {\bibinfo {volume} {9}},\ \bibinfo {pages} {041060} (\bibinfo {year} {2019})},\ \Eprint {https://arxiv.org/abs/1903.08284} {arXiv:1903.08284 [gr-qc]} \BibitemShut {NoStop}%
\bibitem [{\citenamefont {Cheung}\ \emph {et~al.}(2024)\citenamefont {Cheung}, \citenamefont {Berti}, \citenamefont {Baibhav},\ and\ \citenamefont {Cotesta}}]{Cheung:2023vki}%
  \BibitemOpen
  \bibfield  {author} {\bibinfo {author} {\bibfnamefont {M.~H.-Y.}\ \bibnamefont {Cheung}}, \bibinfo {author} {\bibfnamefont {E.}~\bibnamefont {Berti}}, \bibinfo {author} {\bibfnamefont {V.}~\bibnamefont {Baibhav}},\ and\ \bibinfo {author} {\bibfnamefont {R.}~\bibnamefont {Cotesta}},\ }\bibfield  {title} {\bibinfo {title} {{Extracting linear and nonlinear quasinormal modes from black hole merger simulations}},\ }\href {https://doi.org/10.1103/PhysRevD.109.044069} {\bibfield  {journal} {\bibinfo  {journal} {Phys. Rev. D}\ }\textbf {\bibinfo {volume} {109}},\ \bibinfo {pages} {044069} (\bibinfo {year} {2024})},\ \Eprint {https://arxiv.org/abs/2310.04489} {arXiv:2310.04489 [gr-qc]} \BibitemShut {NoStop}%
\bibitem [{\citenamefont {Regge}\ and\ \citenamefont {Wheeler}(1957)}]{Regge:1957td}%
  \BibitemOpen
  \bibfield  {author} {\bibinfo {author} {\bibfnamefont {T.}~\bibnamefont {Regge}}\ and\ \bibinfo {author} {\bibfnamefont {J.~A.}\ \bibnamefont {Wheeler}},\ }\bibfield  {title} {\bibinfo {title} {{Stability of a Schwarzschild singularity}},\ }\href {https://doi.org/10.1103/PhysRev.108.1063} {\bibfield  {journal} {\bibinfo  {journal} {Phys.Rev.}\ }\textbf {\bibinfo {volume} {108}},\ \bibinfo {pages} {1063} (\bibinfo {year} {1957})}\BibitemShut {NoStop}%
\bibitem [{\citenamefont {Zerilli}(1970)}]{Zerilli:1970aa}%
  \BibitemOpen
  \bibfield  {author} {\bibinfo {author} {\bibfnamefont {F.~J.}\ \bibnamefont {Zerilli}},\ }\bibfield  {title} {\bibinfo {title} {Gravitational field of a particle falling in a schwarzschild geometry analyzed in tensor harmonics},\ }\href {https://doi.org/10.1103/PhysRevD.2.2141} {\bibfield  {journal} {\bibinfo  {journal} {Physical Review D}\ }\textbf {\bibinfo {volume} {2}},\ \bibinfo {pages} {2141} (\bibinfo {year} {1970})}\BibitemShut {NoStop}%
\bibitem [{\citenamefont {Ma}\ \emph {et~al.}(2022)\citenamefont {Ma}, \citenamefont {Mitman}, \citenamefont {Sun}, \citenamefont {Deppe}, \citenamefont {H{\'{e}}bert}, \citenamefont {Kidder}, \citenamefont {Moxon}, \citenamefont {Throwe}, \citenamefont {Vu},\ and\ \citenamefont {Chen}}]{Ma_2022}%
  \BibitemOpen
  \bibfield  {author} {\bibinfo {author} {\bibfnamefont {S.}~\bibnamefont {Ma}}, \bibinfo {author} {\bibfnamefont {K.}~\bibnamefont {Mitman}}, \bibinfo {author} {\bibfnamefont {L.}~\bibnamefont {Sun}}, \bibinfo {author} {\bibfnamefont {N.}~\bibnamefont {Deppe}}, \bibinfo {author} {\bibfnamefont {F.}~\bibnamefont {H{\'{e}}bert}}, \bibinfo {author} {\bibfnamefont {L.~E.}\ \bibnamefont {Kidder}}, \bibinfo {author} {\bibfnamefont {J.}~\bibnamefont {Moxon}}, \bibinfo {author} {\bibfnamefont {W.}~\bibnamefont {Throwe}}, \bibinfo {author} {\bibfnamefont {N.~L.}\ \bibnamefont {Vu}},\ and\ \bibinfo {author} {\bibfnamefont {Y.}~\bibnamefont {Chen}},\ }\bibfield  {title} {\bibinfo {title} {Quasinormal-mode filters: A new approach to analyze the gravitational-wave ringdown of binary black-hole mergers},\ }\bibfield  {journal} {\bibinfo  {journal} {Physical Review D}\ }\textbf {\bibinfo {volume} {106}},\ \href {https://doi.org/10.1103/physrevd.106.084036} {10.1103/physrevd.106.084036} (\bibinfo {year} {2022})\BibitemShut
  {NoStop}%
\bibitem [{\citenamefont {London}\ \emph {et~al.}(2014)\citenamefont {London}, \citenamefont {Shoemaker},\ and\ \citenamefont {Healy}}]{London_2014}%
  \BibitemOpen
  \bibfield  {author} {\bibinfo {author} {\bibfnamefont {L.}~\bibnamefont {London}}, \bibinfo {author} {\bibfnamefont {D.}~\bibnamefont {Shoemaker}},\ and\ \bibinfo {author} {\bibfnamefont {J.}~\bibnamefont {Healy}},\ }\bibfield  {title} {\bibinfo {title} {Modeling ringdown: Beyond the fundamental quasinormal modes},\ }\bibfield  {journal} {\bibinfo  {journal} {Physical Review D}\ }\textbf {\bibinfo {volume} {90}},\ \href {https://doi.org/10.1103/physrevd.90.124032} {10.1103/physrevd.90.124032} (\bibinfo {year} {2014})\BibitemShut {NoStop}%
\bibitem [{\citenamefont {Mitman}\ \emph {et~al.}(2023)\citenamefont {Mitman} \emph {et~al.}}]{Mitman:2022qdl}%
  \BibitemOpen
  \bibfield  {author} {\bibinfo {author} {\bibfnamefont {K.}~\bibnamefont {Mitman}} \emph {et~al.},\ }\bibfield  {title} {\bibinfo {title} {{Nonlinearities in Black Hole Ringdowns}},\ }\href {https://doi.org/10.1103/PhysRevLett.130.081402} {\bibfield  {journal} {\bibinfo  {journal} {Phys. Rev. Lett.}\ }\textbf {\bibinfo {volume} {130}},\ \bibinfo {pages} {081402} (\bibinfo {year} {2023})},\ \Eprint {https://arxiv.org/abs/2208.07380} {arXiv:2208.07380 [gr-qc]} \BibitemShut {NoStop}%
\bibitem [{\citenamefont {Cheung}\ \emph {et~al.}(2023)\citenamefont {Cheung} \emph {et~al.}}]{Cheung:2022rbm}%
  \BibitemOpen
  \bibfield  {author} {\bibinfo {author} {\bibfnamefont {M.~H.-Y.}\ \bibnamefont {Cheung}} \emph {et~al.},\ }\bibfield  {title} {\bibinfo {title} {{Nonlinear Effects in Black Hole Ringdown}},\ }\href {https://doi.org/10.1103/PhysRevLett.130.081401} {\bibfield  {journal} {\bibinfo  {journal} {Phys. Rev. Lett.}\ }\textbf {\bibinfo {volume} {130}},\ \bibinfo {pages} {081401} (\bibinfo {year} {2023})},\ \Eprint {https://arxiv.org/abs/2208.07374} {arXiv:2208.07374 [gr-qc]} \BibitemShut {NoStop}%
\bibitem [{\citenamefont {Khera}\ \emph {et~al.}(2023)\citenamefont {Khera}, \citenamefont {Ribes~Metidieri}, \citenamefont {Bonga}, \citenamefont {Jim\'enez~Forteza}, \citenamefont {Krishnan}, \citenamefont {Poisson}, \citenamefont {Pook-Kolb}, \citenamefont {Schnetter},\ and\ \citenamefont {Yang}}]{Khera:2023oyf}%
  \BibitemOpen
  \bibfield  {author} {\bibinfo {author} {\bibfnamefont {N.}~\bibnamefont {Khera}}, \bibinfo {author} {\bibfnamefont {A.}~\bibnamefont {Ribes~Metidieri}}, \bibinfo {author} {\bibfnamefont {B.}~\bibnamefont {Bonga}}, \bibinfo {author} {\bibfnamefont {X.}~\bibnamefont {Jim\'enez~Forteza}}, \bibinfo {author} {\bibfnamefont {B.}~\bibnamefont {Krishnan}}, \bibinfo {author} {\bibfnamefont {E.}~\bibnamefont {Poisson}}, \bibinfo {author} {\bibfnamefont {D.}~\bibnamefont {Pook-Kolb}}, \bibinfo {author} {\bibfnamefont {E.}~\bibnamefont {Schnetter}},\ and\ \bibinfo {author} {\bibfnamefont {H.}~\bibnamefont {Yang}},\ }\bibfield  {title} {\bibinfo {title} {{Nonlinear Ringdown at the Black Hole Horizon}},\ }\href {https://doi.org/10.1103/PhysRevLett.131.231401} {\bibfield  {journal} {\bibinfo  {journal} {Phys. Rev. Lett.}\ }\textbf {\bibinfo {volume} {131}},\ \bibinfo {pages} {231401} (\bibinfo {year} {2023})},\ \Eprint {https://arxiv.org/abs/2306.11142} {arXiv:2306.11142 [gr-qc]} \BibitemShut {NoStop}%
\bibitem [{\citenamefont {Zhu}\ \emph {et~al.}(2024)\citenamefont {Zhu} \emph {et~al.}}]{Zhu:2024rej}%
  \BibitemOpen
  \bibfield  {author} {\bibinfo {author} {\bibfnamefont {H.}~\bibnamefont {Zhu}} \emph {et~al.},\ }\bibfield  {title} {\bibinfo {title} {{Nonlinear Effects In Black Hole Ringdown From Scattering Experiments I: spin and initial data dependence of quadratic mode coupling}},\ }\href@noop {} {\  (\bibinfo {year} {2024})},\ \Eprint {https://arxiv.org/abs/2401.00805} {arXiv:2401.00805 [gr-qc]} \BibitemShut {NoStop}%
\bibitem [{\citenamefont {Redondo-Yuste}\ \emph {et~al.}(2023)\citenamefont {Redondo-Yuste}, \citenamefont {Carullo}, \citenamefont {Ripley}, \citenamefont {Berti},\ and\ \citenamefont {Cardoso}}]{Redondo-Yuste:2023seq}%
  \BibitemOpen
  \bibfield  {author} {\bibinfo {author} {\bibfnamefont {J.}~\bibnamefont {Redondo-Yuste}}, \bibinfo {author} {\bibfnamefont {G.}~\bibnamefont {Carullo}}, \bibinfo {author} {\bibfnamefont {J.~L.}\ \bibnamefont {Ripley}}, \bibinfo {author} {\bibfnamefont {E.}~\bibnamefont {Berti}},\ and\ \bibinfo {author} {\bibfnamefont {V.}~\bibnamefont {Cardoso}},\ }\bibfield  {title} {\bibinfo {title} {{Spin dependence of black hole ringdown nonlinearities}},\ }\href@noop {} {\  (\bibinfo {year} {2023})},\ \Eprint {https://arxiv.org/abs/2308.14796} {arXiv:2308.14796 [gr-qc]} \BibitemShut {NoStop}%
\bibitem [{\citenamefont {Lagos}\ and\ \citenamefont {Hui}(2023)}]{Lagos_2023}%
  \BibitemOpen
  \bibfield  {author} {\bibinfo {author} {\bibfnamefont {M.}~\bibnamefont {Lagos}}\ and\ \bibinfo {author} {\bibfnamefont {L.}~\bibnamefont {Hui}},\ }\bibfield  {title} {\bibinfo {title} {Generation and propagation of nonlinear quasinormal modes of a schwarzschild black hole},\ }\bibfield  {journal} {\bibinfo  {journal} {Physical Review D}\ }\textbf {\bibinfo {volume} {107}},\ \href {https://doi.org/10.1103/physrevd.107.044040} {10.1103/physrevd.107.044040} (\bibinfo {year} {2023})\BibitemShut {NoStop}%
\bibitem [{\citenamefont {Nakano}\ and\ \citenamefont {Ioka}(2007)}]{Nakano:2007cj}%
  \BibitemOpen
  \bibfield  {author} {\bibinfo {author} {\bibfnamefont {H.}~\bibnamefont {Nakano}}\ and\ \bibinfo {author} {\bibfnamefont {K.}~\bibnamefont {Ioka}},\ }\bibfield  {title} {\bibinfo {title} {{Second Order Quasi-Normal Mode of the Schwarzschild Black Hole}},\ }\href {https://doi.org/10.1103/PhysRevD.76.084007} {\bibfield  {journal} {\bibinfo  {journal} {Phys. Rev. D}\ }\textbf {\bibinfo {volume} {76}},\ \bibinfo {pages} {084007} (\bibinfo {year} {2007})},\ \Eprint {https://arxiv.org/abs/0708.0450} {arXiv:0708.0450 [gr-qc]} \BibitemShut {NoStop}%
\bibitem [{\citenamefont {Ioka}\ and\ \citenamefont {Nakano}(2007)}]{Ioka:2007ak}%
  \BibitemOpen
  \bibfield  {author} {\bibinfo {author} {\bibfnamefont {K.}~\bibnamefont {Ioka}}\ and\ \bibinfo {author} {\bibfnamefont {H.}~\bibnamefont {Nakano}},\ }\bibfield  {title} {\bibinfo {title} {{Second and higher-order quasi-normal modes in binary black hole mergers}},\ }\href {https://doi.org/10.1103/PhysRevD.76.061503} {\bibfield  {journal} {\bibinfo  {journal} {Phys. Rev. D}\ }\textbf {\bibinfo {volume} {76}},\ \bibinfo {pages} {061503} (\bibinfo {year} {2007})},\ \Eprint {https://arxiv.org/abs/0704.3467} {arXiv:0704.3467 [astro-ph]} \BibitemShut {NoStop}%
\bibitem [{\citenamefont {Bucciotti}\ \emph {et~al.}(2023)\citenamefont {Bucciotti}, \citenamefont {Kuntz}, \citenamefont {Serra},\ and\ \citenamefont {Trincherini}}]{Bucciotti:2023ets}%
  \BibitemOpen
  \bibfield  {author} {\bibinfo {author} {\bibfnamefont {B.}~\bibnamefont {Bucciotti}}, \bibinfo {author} {\bibfnamefont {A.}~\bibnamefont {Kuntz}}, \bibinfo {author} {\bibfnamefont {F.}~\bibnamefont {Serra}},\ and\ \bibinfo {author} {\bibfnamefont {E.}~\bibnamefont {Trincherini}},\ }\bibfield  {title} {\bibinfo {title} {{Nonlinear quasi-normal modes: uniform approximation}},\ }\href {https://doi.org/10.1007/JHEP12(2023)048} {\bibfield  {journal} {\bibinfo  {journal} {JHEP}\ }\textbf {\bibinfo {volume} {12}},\ \bibinfo {pages} {048}},\ \Eprint {https://arxiv.org/abs/2309.08501} {arXiv:2309.08501 [hep-th]} \BibitemShut {NoStop}%
\bibitem [{\citenamefont {Ma}\ and\ \citenamefont {Yang}(2024)}]{Ma:2024qcv}%
  \BibitemOpen
  \bibfield  {author} {\bibinfo {author} {\bibfnamefont {S.}~\bibnamefont {Ma}}\ and\ \bibinfo {author} {\bibfnamefont {H.}~\bibnamefont {Yang}},\ }\bibfield  {title} {\bibinfo {title} {{The excitation of quadratic quasinormal modes for Kerr black holes}},\ }\href@noop {} {\  (\bibinfo {year} {2024})},\ \Eprint {https://arxiv.org/abs/2401.15516} {arXiv:2401.15516 [gr-qc]} \BibitemShut {NoStop}%
\bibitem [{\citenamefont {Perrone}\ \emph {et~al.}(2024)\citenamefont {Perrone}, \citenamefont {Barreira}, \citenamefont {Kehagias},\ and\ \citenamefont {Riotto}}]{Perrone:2023jzq}%
  \BibitemOpen
  \bibfield  {author} {\bibinfo {author} {\bibfnamefont {D.}~\bibnamefont {Perrone}}, \bibinfo {author} {\bibfnamefont {T.}~\bibnamefont {Barreira}}, \bibinfo {author} {\bibfnamefont {A.}~\bibnamefont {Kehagias}},\ and\ \bibinfo {author} {\bibfnamefont {A.}~\bibnamefont {Riotto}},\ }\bibfield  {title} {\bibinfo {title} {{Non-linear black hole ringdowns: An analytical approach}},\ }\href {https://doi.org/10.1016/j.nuclphysb.2023.116432} {\bibfield  {journal} {\bibinfo  {journal} {Nucl. Phys. B}\ }\textbf {\bibinfo {volume} {999}},\ \bibinfo {pages} {116432} (\bibinfo {year} {2024})},\ \Eprint {https://arxiv.org/abs/2308.15886} {arXiv:2308.15886 [gr-qc]} \BibitemShut {NoStop}%
\bibitem [{\citenamefont {Yi}\ \emph {et~al.}(2024)\citenamefont {Yi}, \citenamefont {Kuntz}, \citenamefont {Barausse}, \citenamefont {Berti}, \citenamefont {Cheung}, \citenamefont {Kritos},\ and\ \citenamefont {Maselli}}]{Yi:2024elj}%
  \BibitemOpen
  \bibfield  {author} {\bibinfo {author} {\bibfnamefont {S.}~\bibnamefont {Yi}}, \bibinfo {author} {\bibfnamefont {A.}~\bibnamefont {Kuntz}}, \bibinfo {author} {\bibfnamefont {E.}~\bibnamefont {Barausse}}, \bibinfo {author} {\bibfnamefont {E.}~\bibnamefont {Berti}}, \bibinfo {author} {\bibfnamefont {M.~H.-Y.}\ \bibnamefont {Cheung}}, \bibinfo {author} {\bibfnamefont {K.}~\bibnamefont {Kritos}},\ and\ \bibinfo {author} {\bibfnamefont {A.}~\bibnamefont {Maselli}},\ }\bibfield  {title} {\bibinfo {title} {{Nonlinear quasinormal mode detectability with next-generation gravitational wave detectors}},\ }\href@noop {} {\  (\bibinfo {year} {2024})},\ \Eprint {https://arxiv.org/abs/2403.09767} {arXiv:2403.09767 [gr-qc]} \BibitemShut {NoStop}%
\bibitem [{\citenamefont {Gleiser}\ \emph {et~al.}(1996)\citenamefont {Gleiser}, \citenamefont {Nicasio}, \citenamefont {Price},\ and\ \citenamefont {Pullin}}]{Gleiser:1995gx}%
  \BibitemOpen
  \bibfield  {author} {\bibinfo {author} {\bibfnamefont {R.~J.}\ \bibnamefont {Gleiser}}, \bibinfo {author} {\bibfnamefont {C.~O.}\ \bibnamefont {Nicasio}}, \bibinfo {author} {\bibfnamefont {R.~H.}\ \bibnamefont {Price}},\ and\ \bibinfo {author} {\bibfnamefont {J.}~\bibnamefont {Pullin}},\ }\bibfield  {title} {\bibinfo {title} {{Second order perturbations of a Schwarzschild black hole}},\ }\href {https://doi.org/10.1088/0264-9381/13/10/001} {\bibfield  {journal} {\bibinfo  {journal} {Class. Quant. Grav.}\ }\textbf {\bibinfo {volume} {13}},\ \bibinfo {pages} {L117} (\bibinfo {year} {1996})},\ \Eprint {https://arxiv.org/abs/gr-qc/9510049} {arXiv:gr-qc/9510049} \BibitemShut {NoStop}%
\bibitem [{\citenamefont {Nicasio}\ \emph {et~al.}(1999)\citenamefont {Nicasio}, \citenamefont {Gleiser}, \citenamefont {Price},\ and\ \citenamefont {Pullin}}]{Nicasio_1999}%
  \BibitemOpen
  \bibfield  {author} {\bibinfo {author} {\bibfnamefont {C.~O.}\ \bibnamefont {Nicasio}}, \bibinfo {author} {\bibfnamefont {R.~J.}\ \bibnamefont {Gleiser}}, \bibinfo {author} {\bibfnamefont {R.~H.}\ \bibnamefont {Price}},\ and\ \bibinfo {author} {\bibfnamefont {J.}~\bibnamefont {Pullin}},\ }\bibfield  {title} {\bibinfo {title} {Collision of boosted black holes: Second order close limit calculations},\ }\bibfield  {journal} {\bibinfo  {journal} {Physical Review D}\ }\textbf {\bibinfo {volume} {59}},\ \href {https://doi.org/10.1103/physrevd.59.044024} {10.1103/physrevd.59.044024} (\bibinfo {year} {1999})\BibitemShut {NoStop}%
\bibitem [{\citenamefont {Gleiser}\ \emph {et~al.}(2000)\citenamefont {Gleiser}, \citenamefont {Nicasio}, \citenamefont {Price},\ and\ \citenamefont {Pullin}}]{Gleiser_2000}%
  \BibitemOpen
  \bibfield  {author} {\bibinfo {author} {\bibfnamefont {R.~J.}\ \bibnamefont {Gleiser}}, \bibinfo {author} {\bibfnamefont {C.~O.}\ \bibnamefont {Nicasio}}, \bibinfo {author} {\bibfnamefont {R.~H.}\ \bibnamefont {Price}},\ and\ \bibinfo {author} {\bibfnamefont {J.}~\bibnamefont {Pullin}},\ }\bibfield  {title} {\bibinfo {title} {Gravitational radiation from schwarzschild black holes: the second-order perturbation formalism},\ }\href {https://doi.org/10.1016/s0370-1573(99)00048-4} {\bibfield  {journal} {\bibinfo  {journal} {Physics Reports}\ }\textbf {\bibinfo {volume} {325}},\ \bibinfo {pages} {41} (\bibinfo {year} {2000})}\BibitemShut {NoStop}%
\bibitem [{\citenamefont {Brizuela}\ \emph {et~al.}(2006)\citenamefont {Brizuela}, \citenamefont {Martin-Garcia},\ and\ \citenamefont {Mena~Marugan}}]{Brizuela:2006ne}%
  \BibitemOpen
  \bibfield  {author} {\bibinfo {author} {\bibfnamefont {D.}~\bibnamefont {Brizuela}}, \bibinfo {author} {\bibfnamefont {J.~M.}\ \bibnamefont {Martin-Garcia}},\ and\ \bibinfo {author} {\bibfnamefont {G.~A.}\ \bibnamefont {Mena~Marugan}},\ }\bibfield  {title} {\bibinfo {title} {{Second and higher-order perturbations of a spherical spacetime}},\ }\href {https://doi.org/10.1103/PhysRevD.74.044039} {\bibfield  {journal} {\bibinfo  {journal} {Phys. Rev. D}\ }\textbf {\bibinfo {volume} {74}},\ \bibinfo {pages} {044039} (\bibinfo {year} {2006})},\ \Eprint {https://arxiv.org/abs/gr-qc/0607025} {arXiv:gr-qc/0607025} \BibitemShut {NoStop}%
\bibitem [{\citenamefont {Brizuela}\ \emph {et~al.}(2007)\citenamefont {Brizuela}, \citenamefont {Martin-Garcia},\ and\ \citenamefont {Marugan}}]{Brizuela:2007zza}%
  \BibitemOpen
  \bibfield  {author} {\bibinfo {author} {\bibfnamefont {D.}~\bibnamefont {Brizuela}}, \bibinfo {author} {\bibfnamefont {J.~M.}\ \bibnamefont {Martin-Garcia}},\ and\ \bibinfo {author} {\bibfnamefont {G.~A.~M.}\ \bibnamefont {Marugan}},\ }\bibfield  {title} {\bibinfo {title} {{High-order gauge-invariant perturbations of a spherical spacetime}},\ }\href {https://doi.org/10.1103/PhysRevD.76.024004} {\bibfield  {journal} {\bibinfo  {journal} {Phys. Rev. D}\ }\textbf {\bibinfo {volume} {76}},\ \bibinfo {pages} {024004} (\bibinfo {year} {2007})},\ \Eprint {https://arxiv.org/abs/gr-qc/0703069} {arXiv:gr-qc/0703069} \BibitemShut {NoStop}%
\bibitem [{\citenamefont {Brizuela}\ \emph {et~al.}(2009)\citenamefont {Brizuela}, \citenamefont {Martin-Garcia},\ and\ \citenamefont {Tiglio}}]{Brizuela:2009qd}%
  \BibitemOpen
  \bibfield  {author} {\bibinfo {author} {\bibfnamefont {D.}~\bibnamefont {Brizuela}}, \bibinfo {author} {\bibfnamefont {J.~M.}\ \bibnamefont {Martin-Garcia}},\ and\ \bibinfo {author} {\bibfnamefont {M.}~\bibnamefont {Tiglio}},\ }\bibfield  {title} {\bibinfo {title} {{A Complete gauge-invariant formalism for arbitrary second-order perturbations of a Schwarzschild black hole}},\ }\href {https://doi.org/10.1103/PhysRevD.80.024021} {\bibfield  {journal} {\bibinfo  {journal} {Phys. Rev. D}\ }\textbf {\bibinfo {volume} {80}},\ \bibinfo {pages} {024021} (\bibinfo {year} {2009})},\ \Eprint {https://arxiv.org/abs/0903.1134} {arXiv:0903.1134 [gr-qc]} \BibitemShut {NoStop}%
\bibitem [{\citenamefont {Leaver}(1985)}]{leaver}%
  \BibitemOpen
  \bibfield  {author} {\bibinfo {author} {\bibfnamefont {E.~W.}\ \bibnamefont {Leaver}},\ }\bibfield  {title} {\bibinfo {title} {An analytic representation for the quasi-normal modes of kerr black holes},\ }\href {http://www.jstor.org/stable/2397876} {\bibfield  {journal} {\bibinfo  {journal} {Proceedings of the Royal Society of London. Series A, Mathematical and Physical Sciences}\ }\textbf {\bibinfo {volume} {402}},\ \bibinfo {pages} {285} (\bibinfo {year} {1985})}\BibitemShut {NoStop}%
\bibitem [{csv()}]{csvQuadratic}%
  \BibitemOpen
  \href@noop {} {\bibinfo {title} {\url{https://github.com/akuntz00/QuadraticQNM}}}\BibitemShut {NoStop}%
\bibitem [{\citenamefont {{Sachs}}(1961)}]{1961RSPSA.264..309S}%
  \BibitemOpen
  \bibfield  {author} {\bibinfo {author} {\bibfnamefont {R.}~\bibnamefont {{Sachs}}},\ }\bibfield  {title} {\bibinfo {title} {{Gravitational Waves in General Relativity. VI. The Outgoing Radiation Condition}},\ }\href {https://doi.org/10.1098/rspa.1961.0202} {\bibfield  {journal} {\bibinfo  {journal} {Proceedings of the Royal Society of London Series A}\ }\textbf {\bibinfo {volume} {264}},\ \bibinfo {pages} {309} (\bibinfo {year} {1961})}\BibitemShut {NoStop}%
\bibitem [{\citenamefont {{Sachs}}(1962)}]{1962RSPSA.270..103S}%
  \BibitemOpen
  \bibfield  {author} {\bibinfo {author} {\bibfnamefont {R.~K.}\ \bibnamefont {{Sachs}}},\ }\bibfield  {title} {\bibinfo {title} {{Gravitational Waves in General Relativity. VIII. Waves in Asymptotically Flat Space-Time}},\ }\href {https://doi.org/10.1098/rspa.1962.0206} {\bibfield  {journal} {\bibinfo  {journal} {Proceedings of the Royal Society of London Series A}\ }\textbf {\bibinfo {volume} {270}},\ \bibinfo {pages} {103} (\bibinfo {year} {1962})}\BibitemShut {NoStop}%
\bibitem [{\citenamefont {{Newman}}\ and\ \citenamefont {{Penrose}}(1962)}]{1962JMP.....3..566N}%
  \BibitemOpen
  \bibfield  {author} {\bibinfo {author} {\bibfnamefont {E.}~\bibnamefont {{Newman}}}\ and\ \bibinfo {author} {\bibfnamefont {R.}~\bibnamefont {{Penrose}}},\ }\bibfield  {title} {\bibinfo {title} {{An Approach to Gravitational Radiation by a Method of Spin Coefficients}},\ }\href {https://doi.org/10.1063/1.1724257} {\bibfield  {journal} {\bibinfo  {journal} {Journal of Mathematical Physics}\ }\textbf {\bibinfo {volume} {3}},\ \bibinfo {pages} {566} (\bibinfo {year} {1962})}\BibitemShut {NoStop}%
\bibitem [{\citenamefont {Spiers}\ \emph {et~al.}(2023)\citenamefont {Spiers}, \citenamefont {Pound},\ and\ \citenamefont {Wardell}}]{Spiers:2023mor}%
  \BibitemOpen
  \bibfield  {author} {\bibinfo {author} {\bibfnamefont {A.}~\bibnamefont {Spiers}}, \bibinfo {author} {\bibfnamefont {A.}~\bibnamefont {Pound}},\ and\ \bibinfo {author} {\bibfnamefont {B.}~\bibnamefont {Wardell}},\ }\bibfield  {title} {\bibinfo {title} {{Second-order perturbations of the Schwarzschild spacetime: practical, covariant and gauge-invariant formalisms}},\ }\href@noop {} {\  (\bibinfo {year} {2023})},\ \Eprint {https://arxiv.org/abs/2306.17847} {arXiv:2306.17847 [gr-qc]} \BibitemShut {NoStop}%
\bibitem [{\citenamefont {Isi}\ and\ \citenamefont {Farr}(2021)}]{Isi:2021iql}%
  \BibitemOpen
  \bibfield  {author} {\bibinfo {author} {\bibfnamefont {M.}~\bibnamefont {Isi}}\ and\ \bibinfo {author} {\bibfnamefont {W.~M.}\ \bibnamefont {Farr}},\ }\bibfield  {title} {\bibinfo {title} {{Analyzing black-hole ringdowns}},\ }\href@noop {} {\  (\bibinfo {year} {2021})},\ \Eprint {https://arxiv.org/abs/2107.05609} {arXiv:2107.05609 [gr-qc]} \BibitemShut {NoStop}%
\bibitem [{\citenamefont {Lim}\ \emph {et~al.}(2019)\citenamefont {Lim}, \citenamefont {Khanna}, \citenamefont {Apte},\ and\ \citenamefont {Hughes}}]{Lim:2019xrb}%
  \BibitemOpen
  \bibfield  {author} {\bibinfo {author} {\bibfnamefont {H.}~\bibnamefont {Lim}}, \bibinfo {author} {\bibfnamefont {G.}~\bibnamefont {Khanna}}, \bibinfo {author} {\bibfnamefont {A.}~\bibnamefont {Apte}},\ and\ \bibinfo {author} {\bibfnamefont {S.~A.}\ \bibnamefont {Hughes}},\ }\bibfield  {title} {\bibinfo {title} {{Exciting black hole modes via misaligned coalescences: II. The mode content of late-time coalescence waveforms}},\ }\href {https://doi.org/10.1103/PhysRevD.100.084032} {\bibfield  {journal} {\bibinfo  {journal} {Phys. Rev. D}\ }\textbf {\bibinfo {volume} {100}},\ \bibinfo {pages} {084032} (\bibinfo {year} {2019})},\ \Eprint {https://arxiv.org/abs/1901.05902} {arXiv:1901.05902 [gr-qc]} \BibitemShut {NoStop}%
\bibitem [{\citenamefont {Bucciotti}\ \emph {et~al.}(2024)\citenamefont {Bucciotti}, \citenamefont {Juliano}, \citenamefont {Kuntz},\ and\ \citenamefont {Trincherini}}]{Bucciotti:2024jrv}%
  \BibitemOpen
  \bibfield  {author} {\bibinfo {author} {\bibfnamefont {B.}~\bibnamefont {Bucciotti}}, \bibinfo {author} {\bibfnamefont {L.}~\bibnamefont {Juliano}}, \bibinfo {author} {\bibfnamefont {A.}~\bibnamefont {Kuntz}},\ and\ \bibinfo {author} {\bibfnamefont {E.}~\bibnamefont {Trincherini}},\ }\bibfield  {title} {\bibinfo {title} {{Amplitudes and Polarizations of Quadratic Quasi-Normal Modes for a Schwarzschild Black Hole}},\ }\href@noop {} {\  (\bibinfo {year} {2024})},\ \Eprint {https://arxiv.org/abs/2406.14611} {arXiv:2406.14611 [hep-th]} \BibitemShut {NoStop}%
\bibitem [{\citenamefont {Abbott}\ \emph {et~al.}(2019)\citenamefont {Abbott} \emph {et~al.}}]{LIGOScientific:2018mvr}%
  \BibitemOpen
  \bibfield  {author} {\bibinfo {author} {\bibfnamefont {B.~P.}\ \bibnamefont {Abbott}} \emph {et~al.} (\bibinfo {collaboration} {LIGO Scientific, Virgo}),\ }\bibfield  {title} {\bibinfo {title} {{GWTC-1: A Gravitational-Wave Transient Catalog of Compact Binary Mergers Observed by LIGO and Virgo during the First and Second Observing Runs}},\ }\href {https://doi.org/10.1103/PhysRevX.9.031040} {\bibfield  {journal} {\bibinfo  {journal} {Phys. Rev. X}\ }\textbf {\bibinfo {volume} {9}},\ \bibinfo {pages} {031040} (\bibinfo {year} {2019})},\ \Eprint {https://arxiv.org/abs/1811.12907} {arXiv:1811.12907 [astro-ph.HE]} \BibitemShut {NoStop}%
\bibitem [{\citenamefont {Riva}\ \emph {et~al.}(2023)\citenamefont {Riva}, \citenamefont {Santoni}, \citenamefont {Savi\'c},\ and\ \citenamefont {Vernizzi}}]{Riva:2023rcm}%
  \BibitemOpen
  \bibfield  {author} {\bibinfo {author} {\bibfnamefont {M.~M.}\ \bibnamefont {Riva}}, \bibinfo {author} {\bibfnamefont {L.}~\bibnamefont {Santoni}}, \bibinfo {author} {\bibfnamefont {N.}~\bibnamefont {Savi\'c}},\ and\ \bibinfo {author} {\bibfnamefont {F.}~\bibnamefont {Vernizzi}},\ }\bibfield  {title} {\bibinfo {title} {{Vanishing of Nonlinear Tidal Love Numbers of Schwarzschild Black Holes}},\ }\href@noop {} {\  (\bibinfo {year} {2023})},\ \Eprint {https://arxiv.org/abs/2312.05065} {arXiv:2312.05065 [gr-qc]} \BibitemShut {NoStop}%
\bibitem [{\citenamefont {Borhanian}\ \emph {et~al.}(2020)\citenamefont {Borhanian}, \citenamefont {Arun}, \citenamefont {Pfeiffer},\ and\ \citenamefont {Sathyaprakash}}]{Borhanian:2019kxt}%
  \BibitemOpen
  \bibfield  {author} {\bibinfo {author} {\bibfnamefont {S.}~\bibnamefont {Borhanian}}, \bibinfo {author} {\bibfnamefont {K.~G.}\ \bibnamefont {Arun}}, \bibinfo {author} {\bibfnamefont {H.~P.}\ \bibnamefont {Pfeiffer}},\ and\ \bibinfo {author} {\bibfnamefont {B.~S.}\ \bibnamefont {Sathyaprakash}},\ }\bibfield  {title} {\bibinfo {title} {{Comparison of post-Newtonian mode amplitudes with numerical relativity simulations of binary black holes}},\ }\href {https://doi.org/10.1088/1361-6382/ab6a21} {\bibfield  {journal} {\bibinfo  {journal} {Class. Quant. Grav.}\ }\textbf {\bibinfo {volume} {37}},\ \bibinfo {pages} {065006} (\bibinfo {year} {2020})},\ \Eprint {https://arxiv.org/abs/1901.08516} {arXiv:1901.08516 [gr-qc]} \BibitemShut {NoStop}%
\bibitem [{\citenamefont {Bourg}\ \emph {et~al.}(2024)\citenamefont {Bourg}, \citenamefont {Panosso~Macedo}, \citenamefont {Spiers}, \citenamefont {Leather}, \citenamefont {Bonga},\ and\ \citenamefont {Pound}}]{Bourg:2024jme}%
  \BibitemOpen
  \bibfield  {author} {\bibinfo {author} {\bibfnamefont {P.}~\bibnamefont {Bourg}}, \bibinfo {author} {\bibfnamefont {R.}~\bibnamefont {Panosso~Macedo}}, \bibinfo {author} {\bibfnamefont {A.}~\bibnamefont {Spiers}}, \bibinfo {author} {\bibfnamefont {B.}~\bibnamefont {Leather}}, \bibinfo {author} {\bibfnamefont {B.}~\bibnamefont {Bonga}},\ and\ \bibinfo {author} {\bibfnamefont {A.}~\bibnamefont {Pound}},\ }\bibfield  {title} {\bibinfo {title} {{Quadratic quasi-normal mode dependence on linear mode parity}},\ }\href@noop {} {\  (\bibinfo {year} {2024})},\ \Eprint {https://arxiv.org/abs/2405.10270} {arXiv:2405.10270 [gr-qc]} \BibitemShut {NoStop}%
\bibitem [{BHP()}]{BHPToolkit}%
  \BibitemOpen
  \href@noop {} {\bibinfo {title} {{Black Hole Perturbation Toolkit}}},\ \bibinfo {howpublished} {(\href{http://bhptoolkit.org/}{bhptoolkit.org})}\BibitemShut {NoStop}%
\end{thebibliography}%

\end{document}